\documentclass[a4paper,11pt]{article}
\pdfoutput=1

\usepackage{jcappub}

\usepackage{amsmath,amssymb,graphicx}
\usepackage{soul}
\usepackage[latin1]{inputenc}
\usepackage{dsfont}

\usepackage{subfigure}

\usepackage{array}
\usepackage{mathtools}
\usepackage{cancel}

\usepackage[thinlines]{easytable}

\usepackage{color}

\usepackage[numbers,sort&compress]{natbib}
\AtBeginDocument{}

\usepackage{multirow}

\usepackage{verbatim}

\usepackage{filecontents}

\newcommand{\be}{\begin{eqnarray}}
\newcommand{\en}{\end{eqnarray}}

\newcommand{\lv}{\langle}
\newcommand{\rv}{\rangle}
\newcommand{\bepm}{\begingroup\setlength{\arraycolsep}{8pt} \renewcommand{\arraystretch}{1} \begin{pmatrix}}
\newcommand{\enpm}{\end{pmatrix} \endgroup}

\usepackage[section, below, above]{placeins}

\usepackage{afterpage}

\begin{document}

\begin{titlepage}

\vspace*{-15mm}
\vspace*{0.7cm}

\begin{center}

{\Large {\bf Predictions from a flavour GUT model combined \\[1mm]  with a SUSY breaking sector}}\\[8mm]

Stefan Antusch$^{\star\dagger}$\footnote{Email: \texttt{stefan.antusch@unibas.ch}},  
Christian Hohl$^{\star}$\footnote{Email: \texttt{ch.hohl@unibas.ch}}

\end{center}

\vspace*{0.20cm}

\centerline{$^{\star}$ \it
Department of Physics, University of Basel,}
\centerline{\it
Klingelbergstr.\ 82, CH-4056 Basel, Switzerland}

\vspace*{0.4cm}

\centerline{$^{\dagger}$ \it
Max-Planck-Institut f\"ur Physik (Werner-Heisenberg-Institut),}
\centerline{\it
F\"ohringer Ring 6, D-80805 M\"unchen, Germany}

\vspace*{1.2cm}

\begin{abstract}
\noindent We discuss how flavour GUT models in the context of supergravity can be completed with a simple SUSY breaking sector, such that the flavour-dependent (non-universal) soft breaking terms can be calculated. As an example, we discuss a model based on an SU(5) GUT symmetry and $A_4$ family symmetry, plus additional discrete ``shaping symmetries'' and a $\mathbb{Z}_4^\mathrm{R}$ symmetry. We calculate the soft terms and identify the relevant high scale input parameters, and investigate the resulting predictions for the low scale observables, such as flavour violating processes, the sparticle spectrum and the dark matter relic density. 

\end{abstract}
\end{titlepage}


\section{Introduction} \label{sec:introduction}

The Standard Model (SM) of elementary particle physics, combined with the concordance ($\Lambda$CDM) model of cosmology, provides an excellent description of most observations to date. However, it fails to explain the observed masses of the neutrinos and it does not provide a suitable candidate for the dark matter (DM) component of the universe. Furthermore, it is plagued by the gauge hierarchy problem, the instability of the Higgs sector under quantum corrections, and, especially in the flavour sector, it has a large number of parameters which are simply fitted to match observations, without providing an explanation where the flavour structure comes from.

One framework towards resolving these shortcomings of the SM in predictive models are supersymmetric flavour Grand Unified Theories (GUTs), where the forces of the SM are unified in one unifying gauge group and where the flavour structure is generated when a family symmetry $G_\mathrm{F}$ is broken. Supersymmetry (SUSY), or its local version supergravity (SUGRA), provides a mechanism for stabilising the hierarchies as well as promising candidates for the dark matter particle, i.e.\ the neutralino or the gravitino (when R parity is conserved or only very weakly broken). We note that in the context of GUTs, a hierarchy stabilisation mechanism is strongly desirable, and even when the sparticle masses are somewhat above the EW scale, this remaining hierarchy is very small compared to the big hierarchy between the electroweak (EW) scale and the GUT scale. 

Typically, flavour GUT models are focussing on the part $W_{\mathrm{mat}}$ of the superpotential, where the Yukawa matrices (and mass matrices) for the matter sector of the theory are contained, which, after GUT symmetry breaking and evolving the parameters to low energies, gives rise to the flavour sector parameters, i.e.\ the masses and mixings of the SM. In addition, they also have to include the superpotential part $W_{\mathrm{fl}}$, where the spontaneous breaking of the family symmetry $G_\mathrm{F}$ is realized by flavour-Higgs fields, so-called ``flavons''. Furthermore, the complete superpotential may be written as
\begin{align}
W = W_{\mathrm{mat}} + W_{\mathrm{fl}} + W_{\cancel{\mathrm{GUT}}} + W_{\cancel{\mathrm{SUSY}}} \;,
\end{align}
where $W_{\cancel{\mathrm{GUT}}}$ denotes the sector of the theory, where the GUT symmetry gets broken, and $W_{\cancel{\mathrm{SUSY}}}$ the SUSY breaking sector.

Regarding $W_{\mathrm{mat}}$ and $W_{\mathrm{fl}}$, various models have been constructed in the literature (for reviews and recent example models see e.g. \cite{King:2017guk}). It has recently been demonstrated that such models can be combined with suitable GUT symmetry breaking sectors in predictive theory frameworks \cite{Antusch:2014poa} which can also resolve the doublet-triplet (DT) splitting challenge in GUTs. Since models for $W_{\mathrm{mat}}$ and $W_{\mathrm{fl}}$ often rely on R symmetry, it has also been discussed in \cite{Antusch:2014poa} how spontaneous GUT and R symmetry breaking can be realized with discrete R symmetries $\mathbb{Z}^\mathrm{R}_n$.

In this work, we investigate how to combine $W_{\mathrm{mat}}$ and $W_{\mathrm{fl}}$ with a superpotential $W_{\cancel{\mathrm{SUSY}}}$ for SUSY breaking in the framework of supergravity. We find that already simple model extensions can achieve this, and allow to calculate the GUT scale input parameters for the soft SUSY breaking terms and the gravitino mass. For an example model we demonstrate that adding a SUSY breaking sector to a predictive GUT flavour model strongly increases the predictivity. E.g., the model can then also predict ranges for the sparticle masses (cf.\ \cite{Antusch:2015nwi,Antusch:2016nak}), for the DM relic density, and for the various flavour violating processes and precision observables which can be used as indirect searches for new physics.

\section{Combining flavour models with a SUSY breaking sector}

\subsection{Flavon superpotentials}\label{sec:flavon_superpot}
In the following we consider SUSY flavour GUT models, where the flavour structure is generated when a family symmetry $G_\mathrm{F}$ is spontaneously broken by vacuum expectation values (vev) of the scalar components of flavon superfields. Schematically, a typical superpotential which realizes a non-zero vev for a flavon superfield $\theta$ has the form
\begin{align}
\label{eq:flavon_1}
W_\mathrm{fl}^\mathrm{\theta} &= P (\kappa  \theta^n - M^2)\:,
\end{align}
where $M$ is a mass scale and $P$ is a so-called driving superfield. In order to break the family symmetry, $\theta$ must transform in a non-trivial representation of $G_\mathrm{F}$, whereas $\theta^n$ is a singlet with respect to this symmetry. The parameter $\kappa$ has mass dimension $-(n-2)$ and is of order $1/\lambda^{n-2}$, with a mass scale $\lambda > M$. Natural units with reduced Planck mass $M_{\mathrm{Pl}} = 1$ are used to simplify notation. With the generalized $F$-terms
\begin{align}
\label{eq:f_term_driving}
F_P &= \kappa \theta^n - M^2 + (\partial_P K) W_\mathrm{fl}^\mathrm{\theta} \;,\\
F_\theta &= n \kappa P \theta^{n-1} + (\partial_\theta K) W_\mathrm{fl}^\mathrm{\theta} \;,
\end{align}
the scalar potential reads
\begin{align}
V(P,\theta) &= e^K (F^i (K^{-1})_i^j F_j^* - 3 |W_\mathrm{fl}^\mathrm{\theta}|^2) \;,
\end{align}
where $K$ and $(K^{-1})_i^j$ are the K\"ahler potential and the inverse of the K\"ahler metric, respectively, and $i,j \in \{P,\theta\}$. For any K\"ahler potential there is a local minimum at $P_{\mathrm{min}}=0$ and $\theta_{\mathrm{min}}=M^{2/n}$ which fulfils
\begin{align}
V(P_{\mathrm{min}},\theta_{\mathrm{min}}) &= 0\;.
\end{align}
Note, that the positive definiteness of the K\"ahler metric in the vacuum guarantees that the masses of the fields are positive. The vev of $\theta$ spontaneously breaks the family symmetry $G_\mathrm{F}$. In general, there exists a set of flavon fields, $\{\theta_i\}$, $i=1,\dots,N_\mathrm{F}$. The operators for the matter sector of the theory, i.e.\ for the Yukawa matrices and the right-handed neutrino mass matrix, contain such flavon fields, and the flavour structure is generated after this breaking.

Two comments are in order:
Firstly, the form of the above superpotential can be enforced/protected by demanding a $U(1)_\mathrm{R}$ symmetry, where $P$ has charge $2$ and $\theta$ is uncharged, as well as a $\mathbb{Z}_n$ symmetry, where $\theta$ has charge $1$ and $P$ charge $0$. As has been discussed in \cite{Antusch:2014poa}, for combining the breaking of an R symmetry with GUT symmetry breaking, it is useful to consider a $\mathbb{Z}_n^\mathrm{R}$ symmetry instead of the $U(1)_\mathrm{R}$ symmetry. This implies that additional terms $P^{n+1} + \dots$ (or $P^{n/2+1} + \dots$ if $n$ is even) are allowed in $W_\mathrm{fl}^\mathrm{\theta}$. In the presence of these extra terms, as discussed in the appendix of \cite{Antusch:2014poa}, additional minima with $P \not= 0$ appear, however the minimum with $P = 0$ still exists and can be used for flavour model building.  

Secondly, one may assume that the fundamental theory is CP symmetric, and that CP violation arises only after symmetry breaking. With such ``spontaneous CP violation'', where $M$ and $\kappa$ are real parameters, the phases of the vevs of the flavons can only take discrete values, as discussed in \cite{Antusch:2011sx}. In the example model, which we discuss in Section~\ref{sec:model}, spontaneous CP violation is assumed.

\subsection{SUGRA breaking with one chiral superfield}
\label{sec:sugrabreaking}

A simple superpotential for introducing SUSY breaking in SUGRA is given by 
\begin{align}
\label{eq:susybr_1}
W^\phi_{\cancel{\mathrm{SUSY}}} &= \mu^2 (\phi + \lambda \phi^m) \;,
\end{align}
where $\mu$ has mass dimension $1$ and $\lambda$ mass dimension $1-m$. A $\mathbb{Z}_n^\mathrm{R}$ symmetry constrains $m$ to values like $1+n$ or $1+n/2$ if $n$ is even, assuming that $\phi$ has charge $2$. To start with, let us consider a minimal K\"ahler potential for $\phi$. We will comment below on the effects of including higher order terms. 
With the generalized $F$-term
\begin{align}
F_\phi &= \mu^2(1 + m\lambda\phi^{m-1}) + (\partial_\phi K) W^\phi_{\cancel{\mathrm{SUSY}}} \;,
\end{align}
the scalar potential reads
\begin{align}
V(\phi) &= e^{K} (F_\phi (\partial_\phi \partial_{\phi^*} K)^{-1} F_\phi^* - 3|W^\phi_{\cancel{\mathrm{SUSY}}}|^2) \;.
\end{align}
Generically, the minima do not satisfy $F_\phi = 0$ and thus break SUSY. Furthermore, with a suitable redefinition of $\phi$ the parameter $\lambda$ can be chosen real, i.e. $\phi \rightarrow \exp(i\alpha/(1-m)) \phi$ where $\alpha$ is the phase of $\lambda$. Note, that the fermionic component is eaten in the super-Higgs mechanism, where the gravitino obtains its mass. The two parameters $\mu$ and $\lambda$ can be chosen such that the minimum satisfies the two constraints
\begin{align}
V(\phi_{\mathrm{min}}) &= 0 \;,  \\
e^{K/2}|W^\phi_{\cancel{\mathrm{SUSY}}}| &= m_{3/2} \;,
\end{align}
with a given value $m_{3/2}$ for the gravitino mass. By an appropriate choice of $\lambda$ the minimum of the potential comes to lie at $V(\phi_{\mathrm{min}})=0$, whereas $\mu^2$ rescales $W^\phi_{\cancel{\mathrm{SUSY}}}$ and allows to fix the value of $m_{3/2}$.

There are two higher dimensional operators in the effective non-minimal K\"ahler potential which we like to discuss in more detail: Firstly, terms like $- \gamma_{\phi \phi} (\phi^*\phi)^2 + \dots$, where the dots indicate higher order terms in $\phi^*\phi$, can shift the vev of the SUSY breaking field $\phi$ to smaller values, below $M_{\mathrm{Pl}}$. This is useful in order to interpret the model in an effective field theory framework. Moreover, in the shifted minimum the masses of $\phi_R$ and $\phi_I$ (with $\phi_R := \mbox{Re}(\phi)/\sqrt{2}$ and $\phi_I := \mbox{Im}(\phi)/\sqrt{2}$) can be increased such that $m_{\phi_R},m_{\phi_I} \gg m_{3/2}$. This is desirable since $\phi_R$ and $\phi_I$ are only weakly (gravitationally) coupled to the fields of the SM, they may otherwise spoil big bang nucleosynthesis (BBN) when they are too light and decay too late.     

Furthermore, in the context of scenarios for early universe cosmology, it is often considered problematic that the sgoldstino field components $\phi_R$ and $\phi_I$ can dominate the energy density of the universe at some intermediate stage, when they oscillate around their minima after inflation with comparatively large amplitudes. However, as argued in \cite{Linde:1996cx,Nakayama:2011wqa}, such a ``cosmic moduli problem'' can be avoided by K\"ahler potential terms of the form $- \gamma_{\phi X} (\phi^*\phi)(X^* X)$, where $X$ represents other fields in the theory which dominate the energy density of the universe during the reheating phase after inflation. Due to this term, with effective coupling $\gamma_{\phi X} \gtrsim 10$, the component fields of the sgoldstino move adiabatically to their minima where they only perform oscillations with negligible amplitudes. 

In the following, we assume that such effective terms in the K\"ahler potential are present, and that we have a ``standard cosmology'' scenario for the later stages of the universe. Moreover, we will consider the case that, with R parity conserved, the neutralino has to provide the dominant component of dark matter.

\subsection{General considerations: $N_\mathrm{D}$ driving fields and $N_\mathrm{F}<N_\mathrm{D}$ flavon fields}
\label{sec:DrivingFields}

In the more general case, we may consider $N_\mathrm{D}$ driving fields $P^{\prime}_i$ and $N_\mathrm{F}<N_\mathrm{D}$ flavon fields $\theta_i$, each charged under a separate symmetry group $\mathbb{Z}_{n_i}$. Again, the fields $P^{\prime}_i$ and $\theta_i$ have charge $2$ and $0$, respectively, under a $\mathbb{Z}_n^\mathrm{R}$ symmetry. As we will now discuss, after suitable redefinition of the fields $P^{\prime}_i$ one can view $N_\mathrm{D} - N_\mathrm{F}$ of the driving fields as SUSY breaking fields, while the remaining $N_\mathrm{F}$ of them will serve as driving fields for the $N_\mathrm{F}$ flavon fields. Without loss of generality, one can make the ansatz
\begin{equation}
\renewcommand{\arraystretch}{1.3}
\label{eq:flavon_susybr_1}
\begin{array}{rccrccccccccccr}
W_{\mathrm{fl}+\cancel{\mathrm{SUSY}}} &=&& P^{\prime}_1 \big( & \kappa^{\prime}_{11} {\theta}_1^{n_1} &+& \kappa^{\prime}_{12} {\theta}_2^{n_2} &+& \kappa^{\prime}_{13} {\theta}_3^{n_3} &+& \dots &+& \kappa^{\prime}_{1N_\mathrm{F}} {\theta}_{N_\mathrm{F}}^{n_{N_\mathrm{F}}} &-& M^{\prime 2}_1 \big) \\
&&+& P^{\prime}_2 \big( & \kappa^{\prime}_{21} {\theta}_1^{n_1} &+& \kappa^{\prime}_{22}{\theta}_2^{n_2} &+& \kappa^{\prime}_{23} {\theta}_3^{n_3} &+& \dots &+& \kappa^{\prime}_{2N_\mathrm{F}} {\theta}_{N_\mathrm{F}}^{n_{N_\mathrm{F}}} &-& M^{\prime 2}_2 \big) \\
&&+& P^{\prime}_3 \big( & \kappa^{\prime}_{31} {\theta}_1^{n_1} &+& \kappa^{\prime}_{32}{\theta}_2^{n_2} &+& \kappa^{\prime}_{33} {\theta}_3^{n_3} &+& \dots &+& \kappa^{\prime}_{3N_\mathrm{F}} {\theta}_{N_\mathrm{F}}^{n_{N_\mathrm{F}}} &-& M^{\prime 2}_3 \big) \\
&&+& \dots\quad\quad \\
&&+& P^{\prime}_{N_\mathrm{D}} \big( & \kappa^{\prime}_{N_\mathrm{D} 1} {\theta}_1^{n_1} &+& \kappa^{\prime}_{N_\mathrm{D} 2} {\theta}_2^{n_2} &+& \kappa^{\prime}_{N_\mathrm{D} 3} {\theta}_3^{n_3} &+& \dots &+& \kappa^{\prime}_{N_\mathrm{D} N_\mathrm{F}} {\theta}_{N_\mathrm{F}}^{n_{N_\mathrm{F}}} &-& M^{\prime 2}_{N_\mathrm{D}} \big) \\
&&+& \dots \;,\,\,\quad
\end{array}
\renewcommand{\arraystretch}{1}
\end{equation}
where the dots in the last line indicate possible additional terms which are higher order in the driving fields $P^{\prime}_i$. Redefining the driving fields $P^{\prime}_i$ (by unitary rotation in field space) one can achieve an upper triangular form for the upper $N_\mathrm{F} \times N_\mathrm{F}$ block of the coupling matrix $\kappa^{\prime}$, whereas the remaining $N_\mathrm{D} - N_\mathrm{F}$ driving fields do not couple to the flavons $\theta_i$ anymore\footnote{This redefinition corresponds to the QR decomposition of a complex rectangular matrix, which in our case is the coupling matrix $\kappa^{\prime}$.}. $W_{\mathrm{fl}+\cancel{\mathrm{SUSY}}}$ now reads
\begin{equation}
\label{eq:flavon_susybr_2}
\renewcommand{\arraystretch}{1.3}
\begin{array}{rccrccccccccccr}
W_{\mathrm{fl}+\cancel{\mathrm{SUSY}}} &=&& P_1 \big( &\kappa_{11} {\theta}_1^{n_1} &+& \kappa_{12} {\theta}_2^{n_2} &+& \kappa_{13} {\theta}_3^{n_3} &+& \dots &+& \kappa_{1N_\mathrm{F}} {\theta}_{N_\mathrm{F}}^{n_{N_\mathrm{F}}} &-& {M}_1^2 \big) \\
&& +& P_2 \big( & 0 &+& \kappa_{22} {\theta}_2^{n_2} &+& \kappa_{23} {\theta}_3^{n_3} &+& \dots &+& \kappa_{2N_\mathrm{F}} {\theta}_{N_\mathrm{F}}^{n_{N_\mathrm{F}}} &-& {M}_2^2 \big) \\
&& +& P_3 \big( & 0 &+& 0 &+& \kappa_{33} {\theta}_3^{n_3} &+& \dots &+& \kappa_{3N_\mathrm{F}} {\theta}_{N_\mathrm{F}}^{n_{N_\mathrm{F}}} &-& {M}_3^2 \big) \\
&& +& \dots\quad\quad\quad \\
&& +& P_{N_\mathrm{F}} \big( & 0 &+& 0 &+& 0 &+& \dots &+& \kappa_{{N_\mathrm{F}}{N_\mathrm{F}}} {\theta}_{N_\mathrm{F}}^{n_{N_\mathrm{F}}} &-& {M}_{N_\mathrm{F}}^2 \big) \\
&& +& P_{N_\mathrm{F}+1} \big( & 0 &+& 0 &+& 0 &+& \dots &+& 0 &-& {M}_{N_\mathrm{F}+1}^2 \big) \\
&& +& \dots\quad\quad\quad \\
&& +& P_{N_\mathrm{D}} \big( & 0 &+& 0 &+& 0 &+& \dots &+& 0 &-& {M}_{N_\mathrm{D}}^2 \big) \\
&& +& \dots\;,\,\quad\quad
\end{array}
\renewcommand{\arraystretch}{1}
\end{equation}
where the primes for $P_i, M_i$ and $\kappa_i$ are dropped to indicate that we are in the new basis of the driving fields. Again, the dots in the last line include higher order terms in the driving fields allowed by the $\mathbb{Z}_n^\mathrm{R}$ symmetry, such as $P_{i}^{n+1}$ (or $P_{i}^{n/2+1}$ if $n$ is even), and also higher order terms mixing different $P_i$. Renaming the superfields $P_{i+N_\mathrm{F}}$ with $i=1 , \dots ,N_\mathrm{D}-N_\mathrm{F}$ to $\phi_i$ and the corresponding mass scales $M_{i+N_\mathrm{F}}$ to $\mu_i$, and including the $\lambda_i \phi_i^m$ terms, we can write
\begin{align}
\label{eq:flavon_susybr_3}
\begin{split}
W_{\mathrm{fl}+\cancel{\mathrm{SUSY}}} &= 
P_1 \big(\kappa_{11} {\theta}_1^{n_1} + \kappa_{12} {\theta}_2^{n_2} + \kappa_{13} {\theta}_3^{n_3} + \dots + \kappa_{1N_\mathrm{F}} {\theta}_{N_\mathrm{F}}^{n_{N_\mathrm{F}}} - {M}_1^2 \big) \\
&\quad + P_2 \big(\kappa_{22}{\theta}_2^{n_2} + \kappa_{23} {\theta}_3^{n_3} + \dots + \kappa_{2N_\mathrm{F}} {\theta}_{N_\mathrm{F}}^{n_{N_\mathrm{F}}} - {M}_2^2 \big) \\ 
&\quad + P_3 \big(\kappa_{33} {\theta}_3^{n_3} + \dots + \kappa_{3N_\mathrm{F}} {\theta}_{N_\mathrm{F}}^{n_{N_\mathrm{F}}} - {M}_3^2 \big) \\
&\quad + \dots \\
&\quad + P_{N_\mathrm{F}} \big( \kappa_{N_\mathrm{F} N_\mathrm{F}} {\theta}_{N_\mathrm{F}}^{n_{N_\mathrm{F}}}- {M}_{N_\mathrm{F}}^2 \big) \\
&\quad + \big[\mu_1^2 \phi_1 + \lambda_1 \phi_1^m \big]
+ \dots
+ \big[ \mu_{N_\mathrm{D}-N_\mathrm{F}}^2 \phi_{N_\mathrm{D}-N_\mathrm{F}} + \lambda_{N_\mathrm{D}-N_\mathrm{F}}\phi_{N_\mathrm{D}-N_\mathrm{F}}^m \big] \\
&\quad + \dots \;.
\end{split}
\end{align}
In summary, we find that starting from a general (schematic) superpotential with $N_\mathrm{D}$ driving fields and $N_\mathrm{F}$ flavon fields, we arrive at $N_\mathrm{D} - N_\mathrm{F}$ superpotential contributions suitable for SUSY breaking, as in Eq.~(\ref{eq:susybr_1}), and $N_\mathrm{F}$ superpotential contributions for driving the vevs of the flavon superfields $\theta_i$. Neglecting the coupling terms between $P_i$ and $\phi_i$, the conditions for the vevs of the flavon fields are recovered iteratively. In a first step, the generalized $F$-term of $P_{N_\mathrm{F}}$, as in Eq.~(\ref{eq:f_term_driving}), is set equal to zero what fixes the vev of $\theta_{N_\mathrm{F}}$. In a second step, by using the value of the vev of $\theta_{N_\mathrm{F}}$, the vanishing $F$-term of $P_{N_\mathrm{F}-1}$ fixes the vev of $\theta_{N_\mathrm{F}-1}$. In the same way the value of the vev of $\theta_{N_\mathrm{F}-3}$ is obtained. This procedure continuous until the vev of $\theta_1$ is fixed in a last step.

In other words, after field redefinitions the superpotential $W_{\mathrm{fl}+\cancel{\mathrm{SUSY}}}$ from Eq.~(\ref{eq:flavon_susybr_1}), with terms allowed by a $\mathbb{Z}_n^\mathrm{R}$ symmetry, separates into a generic flavon potential $W_{\mathrm{fl}}$ for $N_\mathrm{F}$ flavons\footnote{A similar discussion for $N_\mathrm{F}$ flavon fields and the same number of driving fields, with $U(1)_\mathrm{R}$ symmetry, can be found in the appendix of \cite{Antusch:2011sx}.}, a generalized SUSY breaking superpotential $W_{\cancel{\mathrm{SUSY}}} $ with $N_\mathrm{D}-N_\mathrm{F}$ SUSY breaking superfields $\phi_i$, and additional coupling terms between the $P_i$ among each other and with the $\phi_i$. In the next subsection we will argue that the SUSY and $G_\mathrm{F}$ breaking minimum of the combined superpotential $W_{\mathrm{fl}+\cancel{\mathrm{SUSY}}}$, in the presence of these additional coupling terms, is qualitatively the same as discussed above.

\subsection{Example: One driving field $P$ and one SUSY breaking field $\phi$}

In oder to investigate SUSY  and $G_{\mathrm{F}}$ breaking with a combined superpotential of the type $W_{\mathrm{fl}+\cancel{\mathrm{SUSY}}}$ introduced in  Eq.~(\ref{eq:flavon_susybr_3}), we consider a $\mathbb{Z}_4^\mathrm{R}$ symmetry and the simplified case $N_\mathrm{F}=1$ and $N_\mathrm{D}=2$, such that
\begin{align}
\label{eq:1flavon2drfields}
W_{\mathrm{fl}+\cancel{\mathrm{SUSY}}} &= 
 \mu^2 (\phi + \lambda \phi^{3}) + P (\kappa {\theta}^n - {M}^2) + a_1 P^3 + a_2 \phi P^2 +a_3 \phi^2 P \;.
\end{align}
Furthermore, the K\"ahler potential is given by
\begin{align}
K &= \tilde{K}(\phi,\phi^*) + P P^* + \theta \theta^* \;,
\end{align}
where $\tilde{K}$ is the K\"ahler potential of the superfield $\phi$ as discussed in Section~\ref{sec:sugrabreaking}. There is no mixing between $\phi$ and $P$ since $K$ is canonically normalized and additional terms suppressed by the Planck scale are neglected.

Let us first discuss the case where the extra coupling terms $a_i$ are zero. With the generalized $F$-terms
\begin{align}
F_\phi &= \mu^2(1 + 3 \lambda\phi^2) + (\partial_\phi K) W_{\mathrm{fl}+\cancel{\mathrm{SUSY}}} \;, \\
F_P &= \kappa \theta^n - M^2 + (\partial_P K) W_{\mathrm{fl}+\cancel{\mathrm{SUSY}}} \;, \\
F_\theta &= n \kappa P \theta^{n-1} + (\partial_\theta K) W_{\mathrm{fl}+\cancel{\mathrm{SUSY}}} \;,
\end{align}
the scalar potential reads
\begin{align}
\label{eq:1flavon2drfields_1}
V(\phi,P,\theta) &= e^K (F^i (K^{-1})_i^j F_j^* - 3 |W_{\mathrm{fl}+\cancel{\mathrm{SUSY}}}|^2) \;,
\end{align}
where $i,j \in \{ \phi,P,\theta \}$. Compared to the flavon superpotential in Eq.~(\ref{eq:flavon_1}), the superpotential $|W|$ (and thus $m_{3/2}$) is no longer zero in the minimum. In the following, it is assumed that family symmetry breaking takes place at scales $M_i \gg m_{3/2}$.

We have numerically studied the scalar potential given in Eq.~(\ref{eq:1flavon2drfields_1}). In summary, we found that the shifts in the minima of the fields, induced by combining the flavon potential with the SUSY breaking potential as in $W_{\mathrm{fl}+\cancel{\mathrm{SUSY}}}$ of Eq.~(\ref{eq:1flavon2drfields}), do not qualitatively change the picture, and a combined solution with spontaneous breaking of the family symmetry $G_{\mathrm{F}}$ and of SUSY is possible in this simple scheme.

This result can also be understood with the following arguments. In order to have a minimum of $V(\phi,P,\theta)$ the two equations $\partial V / \partial P=0$ and $\partial V / \partial \theta=0$ have to be fulfilled. In first approximation this is obtained, if there is a shift of $P$ away from zero by $\mathcal{O}(m_{3/2} \cdot M^{2-n}/\kappa)$, what corresponds to $F_\theta = 0$, and a relative shift of $\theta$ away from $M^2/\kappa$ by $\mathcal{O}(m^2_{3/2}/M^2 \cdot M^{2-n}/\kappa)$, what corresponds to $F_P = 0$ and also compensates for the correction coming from the term $-3|W_{\mathrm{fl}+\cancel{\mathrm{SUSY}}}|^2$. Regarding the minimum for $\phi$, there is only a negligible correction of $\mathcal{O}(m_{3/2}^2)$, since $|W|$ is only changed by a contribution from the flavon sector of $\mathcal{O}(m_{3/2}^3)$. In addition, the parameter $\lambda$ has to be shifted by $\mathcal{O}(m_{3/2}^3)$ in order that the minimum lies at $V=0$.

Let us now discuss the additional coupling terms $a_1 P^3$,  $a_2 \phi P^2$ and $a_3 \phi^2 P$: The first term, as mentioned in Section~\ref{sec:flavon_superpot}, generates another minimum with $P\not=0$, however the minimum with $P \approx 0$ still persists and can be used for model building. The two other terms leave $F_\theta$ unchanged but modify $F_P$ and $F_\phi$. Although the vev of $\phi$ can be large, only somewhat below the Planck scale, the calculations showed that both terms do not qualitatively change the picture, since the additional terms can be absorbed (when plugging in the vev of $\phi $) by a suitable redefinition of $M$ such that the vev of $\theta$ remains unchanged. The only restriction to the parameters $a_i$ is given by $a_3 \langle \phi \rangle^2 \le M^2$, in order that the relative correction to $M$ is of order one or smaller. In contrast, there is a bigger shift in the scalar component of $\phi$ and in the parameter $\lambda$ of $\mathcal{O}(M^2)$. The arguments of this section can also be applied to more general flavon potentials.

\section{An example flavour GUT model with neutrino mixing from CSD2}\label{sec:model}
In this section, we discuss the combination of a flavour GUT model in supergravity, with a SUSY breaking sector along the lines discussed in the previous section. For the flavour GUT model, we closely follow \cite{Antusch:2013wn}, which is based on an SU(5) GUT symmetry and an $A_4$ family symmetry $G_\mathrm{F}$, plus additional discrete ``shaping symmetries''. 

The model breakes CP symmetry spontaneously, via the ``discrete vacuum alignment mechanism'' \cite{Antusch:2011sx}, and explains the right-angled unitarity triangle in the quark sector (where $\alpha \approx 90^\circ$) by realizing the quark phase sum rule \cite{Antusch:2009hq}. The lepton mixing is predicted by the CSD2 scheme \cite{Antusch:2011ic}, plus a charged lepton mixing contribution. 

There are also some changes compared to \cite{Antusch:2013wn}: Most importantly, we consider a $\mathbb{Z}_4^\mathrm{R}$ symmetry instead of a $U(1)_\mathrm{R}$ symmetry, add a simple SUSY breaking sector, and choose somewhat different operators for realizing the Yukawa sector (which now predicts the approximate GUT relations $y_\tau = y_b$ and $y_\mu = -3 y_s$).

\subsection{The flavon potential and the flavon vevs}
For $W_\mathrm{fl}$, we can use the results from \cite{Antusch:2013wn}. As we discussed above, adding the SUSY breaking sector as well as considering the additional coupling terms to the SUSY breaking field(s) only small shifts in the vevs of the driving fields and the flavon fields are induced, suppressed by the small gravitino mass. In addition, because of the $\mathbb{Z}_4^\mathrm{R}$ symmetry instead of the $U(1)_\mathrm{R}$ symmetry, there are also couplings between different driving fields, however their effects are negligible, since all driving field vevs are only $\mathcal{O}(m_{3/2} \cdot M^{2-n}/\kappa)$. 

In the following, we use the notation of \cite{Antusch:2013wn} for the field names, with flavons renamed from $\phi_i$ to $\theta_i$ to match our notation in the previous section. The most relevant flavons for the flavour structure are the $A_4$ triplets $\theta_{23},\theta_{102},\theta_{2}$ and $\theta_{3}$, which have vevs in the following directions in flavour space
\begin{align} \label{eq:FlavonDirectionsNu}
\lv \theta_{23} \rv \sim \begin{pmatrix} 0 \\ 1 \\ -1 \end{pmatrix}, \quad
\lv \theta_{102} \rv \sim \begin{pmatrix} 1 \\ 0 \\  2 \end{pmatrix}, \quad
\lv \theta_{2} \rv \sim \begin{pmatrix} 0 \\ 1 \\  0 \end{pmatrix}, \quad
\lv \theta_{3} \rv \sim \begin{pmatrix} 0 \\ 0 \\  1 \end{pmatrix}.
\end{align}

\subsection{SUSY breaking sector and the matter superpotential} 
\label{sec:SUSYBreakingMatterSuperpotential}
In addition to $W_\mathrm{fl}$, we will consider the simple SUSY breaking sector from Eq.~(\ref{eq:susybr_1})
\begin{align}
W_{\cancel{\mathrm{SUSY}}} &= \mu^2 (\phi + \lambda \phi^n) \;.
\end{align}
The superfield $\phi$ is uncharged under $G_\mathrm{F}$ and only carries charge $2$ under the $\mathbb{Z}_4^\mathrm{R}$ symmetry. The representations and charges under all symmetries of the SU(5) matter multiplets $F$, $T_1$, $T_2$, $T_3$, $N_1$ and $N_2$ as well as of the Higgs and flavon fields are given in Appendix~\ref{secApp:Superpotential}. After integrating out the heavy messenger fields, the superpotential for the matter sector is given by 
\begin{align}
W_\mathrm{mat} &= W_N + W_\nu + W_d + W_u \:,
\end{align}
where the different terms have the form
\begin{align}
\label{eq:SuperpotentialMatterSector}
\begin{split}
W_N &= \xi_1 N_1^2 + \xi_2 N_2^2 \:, \\
W_\nu &= \frac{1}{\Lambda} (H_5 F)(\theta_{23} N_1) + \frac{1}{\Lambda} (H_5 F)(\theta_{102} N_2) \:, \\
W_d &= \frac{1}{\Lambda^3} \theta^{\prime}_2 \bar{H}_5 F (T_1 \theta_2) H_{24} + \frac{1}{\Lambda^3} \theta^{\prime}_{102} \bar{H}_5 F (T_2 \theta_{102}) H_{24} + \frac{1}{\Lambda^2} F (T_2 \theta_{23}) \bar{H}^{\prime}_{5} H_{24} + \frac{1}{\Lambda} \bar{H}_5 F (T_3 \theta_3) \:, \\
W_u &= \frac{1}{\Lambda^2} T_1^2 H_5 \xi_u \xi_1 + \frac{1}{\Lambda^2} T_1 T_2 H_5 \xi_u^2 + \frac{1}{\Lambda^2} T_2^2 H_5 \xi_1^2 + \frac{1}{\Lambda} T_2 T_3 H_5 \xi_1 + T_3^2 H_5 \:,
\end{split}
\end{align}
with the messenger scale $\Lambda$. At the GUT scale they lead to the Yukawa matrices
\begin{equation}
\label{eq:YukawaMatrices}
\begin{array}{rclrcl}
Y_d &=& 
\bepm
0 & \bar{\omega} \epsilon_{102} & 0 \\
i \epsilon_2 & \epsilon_{23} & 0 \\
0 & 2 \bar{\omega} \epsilon_{102} - \epsilon_{23} & \epsilon_3
\enpm , \; &
Y_u &=& 
\bepm
a_u & b_u & 0 \\
b_u & c_u & d_u \\
0 & d_u & e_u \\
\enpm ,
\\\\
Y_e &=& 
\bepm
0 & i \epsilon_2 & 0 \\
\bar{\omega} \epsilon_{102} & -3 \epsilon_{23} & 2 \bar{\omega} \epsilon_{102} + 3 \epsilon_{23} \\
0 & 0 & \epsilon_3 \\
\enpm , \; &
Y_\nu &=& 
\bepm
0 & a & -a \\
b & 0 & 2b
\enpm ,
\end{array}
\end{equation}
as well as to the right-handed neutrino mass matrix
\begin{equation}
\label{eq:MassMatrixRightHandedNeutrinos}
\begin{array}{rcl}
M_R &=& 
\bepm
M_A & 0 \\
0 & M_B
\enpm .
\end{array}
\end{equation}
We followed the notation used in SusyTC \cite{Antusch:2015nwi}, which in particular is the RL convention for the definition of the Yukawa matrices. The $\theta^{\prime}_2$, $\theta^{\prime}_{102}$, $\xi_1$, $\xi_2$ and $\xi_u$ are additional flavon fields that are singlets under $A_4$. Whilst the vevs of  $\xi_1$ and $\xi_u$ are real, the ones of $\xi_2$, $\theta_2^{\prime}$ and $\theta_{102}^{\prime}$ have a complex phase of $-\pi/3$, $\pi/2$ and $4\pi/3$, respectively. The parameters in $Y_d$ and $Y_e$ are defined as
\begin{align}
\epsilon_{23} \sim \frac{v_{24}}{\Lambda^2}|\lv\theta_{23}\rv|,\: \epsilon_{102} \sim \frac{v_{24}}{\Lambda^3}|\lv\theta^{\prime}_{102}\rv \lv\theta_{102}\rv|,\: \epsilon_{2} \sim \frac{v_{24}}{\Lambda^3}|\lv\theta^{\prime}_2\rv \lv\theta_2\rv|,\: \epsilon_{3} \sim \frac{1}{\Lambda}|\lv\theta_3\rv| \:,
\end{align}
where $v_{24}$ is the vev of $H_{24}$, and where the phase $\bar{\omega}$, which corresponds to the phase of $\langle \theta^{\prime}_{102} \rangle$, has the value $\mathrm{e}^{4\pi i/3}$. In addition, the parameters in $Y_u$ are given by
\begin{align}
a_u \sim \frac{|\lv\xi_u\rv \lv\xi_1\rv|}{\Lambda^2},\: b_u \sim \frac{|\lv\xi_u\rv|^2}{\Lambda^2},\: c_u \sim \frac{|\lv\xi_1\rv|^2}{\Lambda^2},\: d_u \sim \frac{|\lv\xi_u\rv|}{\Lambda}\:,
\end{align}
whereas $e_u$ is just coming from a renormalizable coupling. In $Y_\nu$ and $M_R$ the parameters $a$, $b$ and $M_A$ are real and $M_B$ has a complex phase of $-\pi/3$, what corresponds to the phase of $\langle \xi_2 \rangle$. The mass matrix of the light neutrinos follows from the seesaw formula \cite{seesaw}
\begin{align}
\label{eq:SeeSaw}
m_\nu &= \frac{v^2_u}{2} Y_\nu^\top M_R^{-1} Y_\nu \;,
\end{align}
where $v_u = v \sin \beta$ and $v$ is the SM-like EW Higgs vev. Inserting Eq.~(\ref{eq:YukawaMatrices}), (\ref{eq:MassMatrixRightHandedNeutrinos}), we obtain
\begin{equation}
\begin{array}{rcl}
m_\nu &=& \frac{v^2_u}{2}
\bepm
B & 0 & 2B \\
0 & A & -A \\
2B & -A & A+4B
\enpm \;,\quad
\text{with} \quad A = \frac{a^2}{M_A} ,\; B = \frac{b^2}{M_B} \;.
\end{array}
\end{equation}
Since $m_\nu$ only depends on the ratios $a^2/M_A$ and $b^2/M_B$, we are free to fix two of the four parameters which enter $Y_\nu$ and $M_R$.

We have checked that due to the change from $U(1)_\mathrm{R}$ symmetry to $\mathbb{Z}_4^\mathrm{R}$, no dangerous terms are generated. The only new terms at the renormalizable level are some specific trilinear couplings between messenger fields, which however only generate suppressed higher order operators and do not affect the model predictions.

\subsection{Soft SUSY breaking terms}
The soft SUSY breaking terms emerge from the scalar potential after the field in the SUSY breaking sector acquired its vev. Since the vev is chosen close to the Planck scale, the Lagrangian at the GUT scale is considered in the flat limit, where $M_\mathrm{Pl} \rightarrow \infty$ and $m_{3/2}$ is kept fixed. In this limit the SUSY breaking sector decouples from the flavour sector and the Lagrangian has the same form as in global SUSY augmented by the soft SUSY breaking terms. The soft terms are naturally located at the scale of the gravitino mass. Due to the symmetry properties of the fields the scalar trilinear coupling matrices of the squarks and the sleptons have the same structure as the Yukawa matrices in Eq.~(\ref{eq:YukawaMatrices}), up to different factors $k_i$ with mass dimension one in the non-zero entries. These extra factors arise, since in general the parameters in front of each term in the superpotential (Eq.~(\ref{eq:SuperpotentialMatterSector})) and in the K\"ahler potential are actually functions of the SUSY breaking field. More explicitly, in addition to each operator in Eq.~(\ref{eq:SuperpotentialMatterSector}) there exist additional operators with e.g. an extra factor $\phi^2/M^2_{\mathrm{Pl}}$ times an order one parameter, which yields the leading, non-universal contribution to the scalar trilinear coupling matrices. Thus, the $k_i$ can be treated as free parameters of order $m_{3/2}$ and each of them corresponds to one term in the superpotential. Because we assume spontaneous CP violation, the $k_i$ are real. Taking these considerations into account, the scalar trilinear coupling matrices have the form
\begin{equation}
\label{eq:SoftTrilinearMatrices}
\begin{array}{rclrcl}
T_d &=& 
\bepm
0 & k_2 \bar{\omega} \epsilon_{102} & 0 \\
i k_1 \epsilon_2 & k_3 \epsilon_{23} & 0 \\
0 & 2 k_2 \bar{\omega} \epsilon_{102} - k_3 \epsilon_{23} & k_4 \epsilon_3
\enpm , \; &
T_u &=& 
\bepm
k_5 a_u & k_6 b_u & 0 \\
k_6 b_u & k_7 c_u & k_8 d_u \\
0 & k_8 d_u & k_9 e_u \\
\enpm ,
\\\\
T_e &=& 
\bepm
0 & i k_1 \epsilon_2 & 0 \\
k_2 \bar{\omega} \epsilon_{102} & -3 k_3 \epsilon_{23} & 2 k_2 \bar{\omega} \epsilon_{102} + 3 k_4 \epsilon_{23} \\
0 & 0 & \epsilon_3 \\
\enpm , \; &
T_\nu &=& 
\bepm
0 & k_{10} a & -k_{10}a \\
k_{11} b & 0 & 2 k_{11} b
\enpm .
\end{array}
\end{equation}
Since the three matter 5-plets $F_i$ are embedded into a triplet of $A_4$, the corresponding soft scalar mass matrix is proportional to the identity matrix, neglecting subleading corrections from flavon vevs, which are suppressed by the Planck scale. In constrast to that, the mass matrix of the three matter 10-plets $T_i$ has a diagonal form too, but in general the entries on the diagonal are not all the same. According to the embedding of the quarks and the leptons into the 5- and the 10-plets, the squared soft scalar mass matrices are given by
\begin{equation}
\label{eq:SoftScalarMassMatrices_1}
m^2_L = m^2_d =
\bepm
m^2_F & 0 & 0 \\
0 & m^2_F & 0 \\
0 & 0 & m^2_F
\enpm ,\quad
m^2_Q = m^2_u = m^2_e =
\bepm
m^2_{T1} & 0 & 0 \\
0 & m^2_{T2} & 0 \\
0 & 0 & m^2_{T3}
\enpm ,
\end{equation}
where $m^2_F$ and $m^2_{Ti}$ are the squared soft masses of $F$ and $T_i$. Again, they can be treated as free parameters of order $m_{3/2}^2$. In addition, the squared soft masses of the right-handed neutrinos have the form
\begin{equation}
\label{eq:SoftScalarMassMatrices_2}
m^2_\nu = 
\bepm
m^2_{\nu1} & 0 \\
0 & m^2_{\nu2}
\enpm ,
\end{equation}
and the ones of the two MSSM Higgs doublets are denoted by $m^2_{H_d}$ and $m^2_{H_u}$, respectively. Since SU(5) is a simple Lie group, and assuming a simple gauge kinetic function of the form $f_{ab} = f(\phi) \delta_{ab}$, the three soft masses of the MSSM's gauginos are all given by the same mass parameter $m_{\lambda}$ at the GUT scale. For the definition of the soft terms we followed again the notation used in SusyTC \cite{Antusch:2015nwi}.

\section{MCMC analysis and predictions of the example flavour GUT model}\label{sec:analysis}
The Yukawa matrices and the mass matrix of the right-handed neutrinos were presented at the GUT scale $M_{\text{GUT}} = 2 \cdot 10^{16}\,\text{GeV}$. By taking into account spontaneous SUSY breaking in the SUSY breaking sector, the soft scalar trilinear couplings and the soft scalar masses of the squarks and the sleptons are determined, too. However, in order to compare predictions of this model with the experimental data, the corresponding values at low energies, for instance at the mass scale of the $Z$ boson $m_Z \simeq 91\,\text{GeV}$, have to be calculated. Beside the renormalization group (RG) running from $M_{\text{GUT}}$ to $m_Z$, threshold corrections of the heavy superpartners also have to be taken into account, when matching the MSSM to the SM at the SUSY scale $\Lambda_{\text{SUSY}}$.

\subsection{Numerical procedure}
The numerical analysis is performed in the following way: Using the one-loop MSSM and soft term RGEs, we run the parameters from $M_{\text{GUT}} = 2 \cdot 10^{16}\,\text{GeV}$ down to $\Lambda_\mathrm{SUSY}$ with the Mathematica package SusyTC \cite{Antusch:2015nwi}, which is an extension of the Mathematica package REAP \cite{Antusch:2005gp}. The heavy, right-handed neutrinos are integrated out at their corresponding mass scales. The SUSY scale is determined dynamically by the geometric mean of the stop masses $\Lambda_{\mathrm{SUSY}} = \sqrt{m_{\tilde{t}_1}m_{\tilde{t}_2}}$, where the stop masses are defined by the up-type squark mass eigenstates $\tilde{u}_i$ with largest mixing to $\tilde{t}_1$ and $\tilde{t}_2$. The parameters of the model at this mass scale are used to calculate the mass of the SM-like EW Higgs with FeynHiggs \cite{Bahl:2016brp,Hahn:2013ria,Frank:2006yh,Degrassi:2002fi,Heinemeyer:1998np,Heinemeyer:1998yj}, the properties of dark matter with MicrOMEGAs \cite{Belanger:2010pz} and the observables related to flavour violating processes with SUSY FLAVOR \cite{Rosiek:2014sia,Crivellin:2012jv,Rosiek:2010ug}. All superpartners of the SM particles are integrated out at the SUSY scale and the MSSM is matched to the SM at this stage. Finally, we evolve the Yukawa matrices from $\Lambda_{\text{SUSY}}$ to $M_Z = 91.2\,\text{GeV}$ using the one-loop SM REGs in SusyTC and calculate the Yukawa couplings of the quarks and the charged leptons, the masses of the light, left-handed neutrinos and the CKM and PMNS matrix. The masses for the left-handed neutrinos are obtained from the seesaw formula (Eq.~(\ref{eq:SeeSaw})). We choose to fix the masses of the right-handed neutrinos in Eq.~(\ref{eq:MassMatrixRightHandedNeutrinos}) as $|M_A| = 2 \cdot 10^{10}\,\text{GeV}$ and $|M_B| = 2 \cdot 10^{11}\,\text{GeV}$, such that the mass matrix of the light neutrinos only depends on $a$ and $b$. Note, that as long as we are in the regime where the neutrino Yukawa couplings are $\ll1$, the choice of the right-handed neutrino masses (to a good approximation) only affects the values of the parameters $a$ and $b$. Furthermore, the sign of the $\mu$-term is chosen negative and we assume normal ordering of the light neutrino masses.
\\\\
Our model contains $30$ parameters at the GUT scale: There are $12$ parameters from the MSSM, $\epsilon_2$, $\epsilon_{102}$, $\epsilon_{23}$, $\epsilon_3$, $a_u$, $b_u$, $c_u$, $d_u$, $e_u$, $a$, $b$, $\tan\beta$ (Eq.~(\ref{eq:YukawaMatrices})), and $18$ parameters from the soft terms, i.e. from the scalar trilinear terms, $k_i$ with $i=1,\dots,11$, (Eq.~(\ref{eq:SoftTrilinearMatrices})), from the scalar masses, $m_F$, $m_{T_1}$, $m_{T_2}$, $m_{T_3}$, $m_{H_u}$, $m_{H_d}$ (Eq.~(\ref{eq:SoftScalarMassMatrices_1}), (\ref{eq:SoftScalarMassMatrices_2})), and the gaugino mass $m_\lambda$. The Markov chain Monte Carlo (MCMC) analysis will show that the soft scalar trilinear terms at the GUT scale are either restricted to absolute values smaller than $10^4\,\text{GeV}$ by the observables or that their value has no big impact on the low energy parameters. Therefore, in order to avoid a short lifetime of the metastable EW vacuum, what can be caused by big soft scalar trilinear terms, we implement a prior to restrict the parameters $k_i$ to values between $-10^4\,\text{GeV}$ and $10^4\,\text{GeV}$. Since per definition the GUT scale input values of the soft scalar masses are non-negative, a prior is implemented to ensure this. For all parameters the distribution of the corresponding prior is chosen flat.
\\\\
In order to fit the parameters in the model we use the following $30$ observables: We have the Yukawa couplings of the quarks and the charged leptons, $y_u$, $y_c$, $y_t$, $y_d$, $y_s$, $y_b$, $y_e$, $y_\mu$, $y_\tau$, and the CKM parameters, $\theta^{\text{CKM}}_{12}$, $\theta^{\text{CKM}}_{23}$, $\theta^{\text{CKM}}_{13}$, $\delta^{\text{CKM}}$, at $M_Z$ given in \cite{Antusch:2013jca}. Although the Yukawa couplings are given there with high precision, we set their uncertainty to one percent, what is roughly in accordance with the accuracy of the running used here. Furthermore, there are the PMNS parameters $\sin^2(\theta_{12}^{\text{PMNS}})$, $\sin^2(\theta_{23}^{\text{PMNS}})$, $\sin^2(\theta_{13}^{\text{PMNS}})$, and the squared mass differences of the light neutrinos $\Delta m^2_\mathrm{sol}$, $\Delta m^2_\mathrm{atm}$, taken from \cite{Esteban:2016qun,nufit}, and the dark matter relic density $\Omega$ and the SM-like EW Higgs mass $m_h$, given in \cite{Olive:2016xmw}. For the Higgs mass we take an error of $\pm 3\text{GeV}$, what is roughly the uncertainty of the theoretical calculation. The branching ratios of the flavour violating processes $\mu \rightarrow e\gamma$, $\tau \rightarrow e\gamma$, $\tau \rightarrow \mu\gamma$, $K^0_L \rightarrow \pi^0\bar{\nu}\nu$, $K^+ \rightarrow \pi^+\bar{\nu}\nu$, $B^0_S \rightarrow e^+e^-$, $B^0_S \rightarrow \mu^+\mu^-$, $B^0_S \rightarrow e^\pm\mu^\mp$, $B \rightarrow \tau\nu$ and $\epsilon_K$, which indicates the CP violation in the $K^0-\bar{K}^0$ mixing, are taken from \cite{Olive:2016xmw}. Since there is a big uncertainty in the theoretical calculation of $\epsilon_K$, for our analysis we consider the ratio $\epsilon_K^{\text{exp}}/\epsilon^{\text{SM}}_K$, where $\epsilon^{\text{SM}}_K$ is the value in the SM calculated by SUSY FLAVOR.
\\\\
We perform a MCMC analysis, using a Metropolis-Hastings algorithm, to fit the parameters to the measured observables and to calculate the posterior density of the parameters and of the observables. Since in the space of the soft scalar trilinear parameters $k_4$ and $k_9$ the $\chi^2$-function of the dark matter relic density $\Omega$ has several local minima, which are separated by regions with bigger $\chi^2$, for the MCMC analysis we choose ten times the experimental error of $\Omega$, in order to better resolve the whole region in parameter space with a suitable $\chi^2$. The range of $k_4$ and $k_9$ remains the same as in case of considering the experimental error of $\Omega$, because beyond the region of these local minima the $\chi^2$ increases very rapidly and the other observables do not vary much within this region.

\subsection{Results}
Following the procedure described above, the $1\sigma$ highest posterior density (HPD) intervals and the mode values are determined for each parameter. The result is presented in Table~\ref{tab:BestFitParameters}. The smallest $\chi^2$ in the MCMC analysis lie around $20$. If there are UFB directions or CCB minima, we checked that the lifetime of the metastable EW vacuum is much bigger than the age of the universe, following \cite{Casas:1995pd,Bajc:2015ita,Sarid:1998sn,Coleman:1977py}. The $1\sigma$ HPD intervals and the mode values for the observables are shown in Table~\ref{tab:BestFitObservables}.

\begin{figure}
\begin{center}
\includegraphics[width=0.9\textwidth]{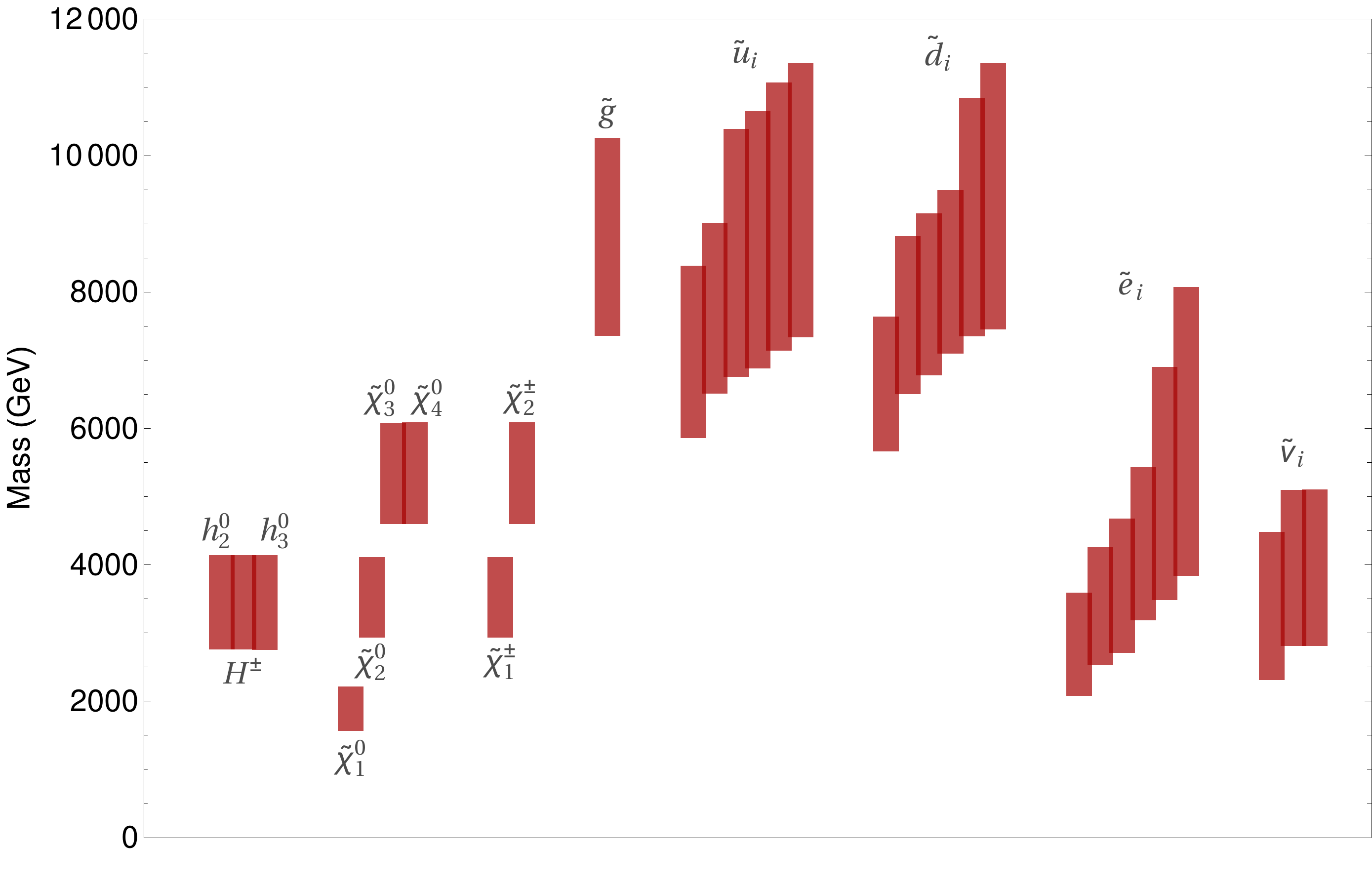}
\end{center}
\caption{$1\sigma$ HPD intervals of the sparticle and the heavy Higgs boson masses. The LSP is always the neutralino $\tilde{\chi}^0_1$ and the NLSP is either a chargino, a stau or a sneutrino.}
\label{fig:MassSpectrum}
\end{figure}

\begin{figure}
\begin{center}
\includegraphics[width=0.9\textwidth]{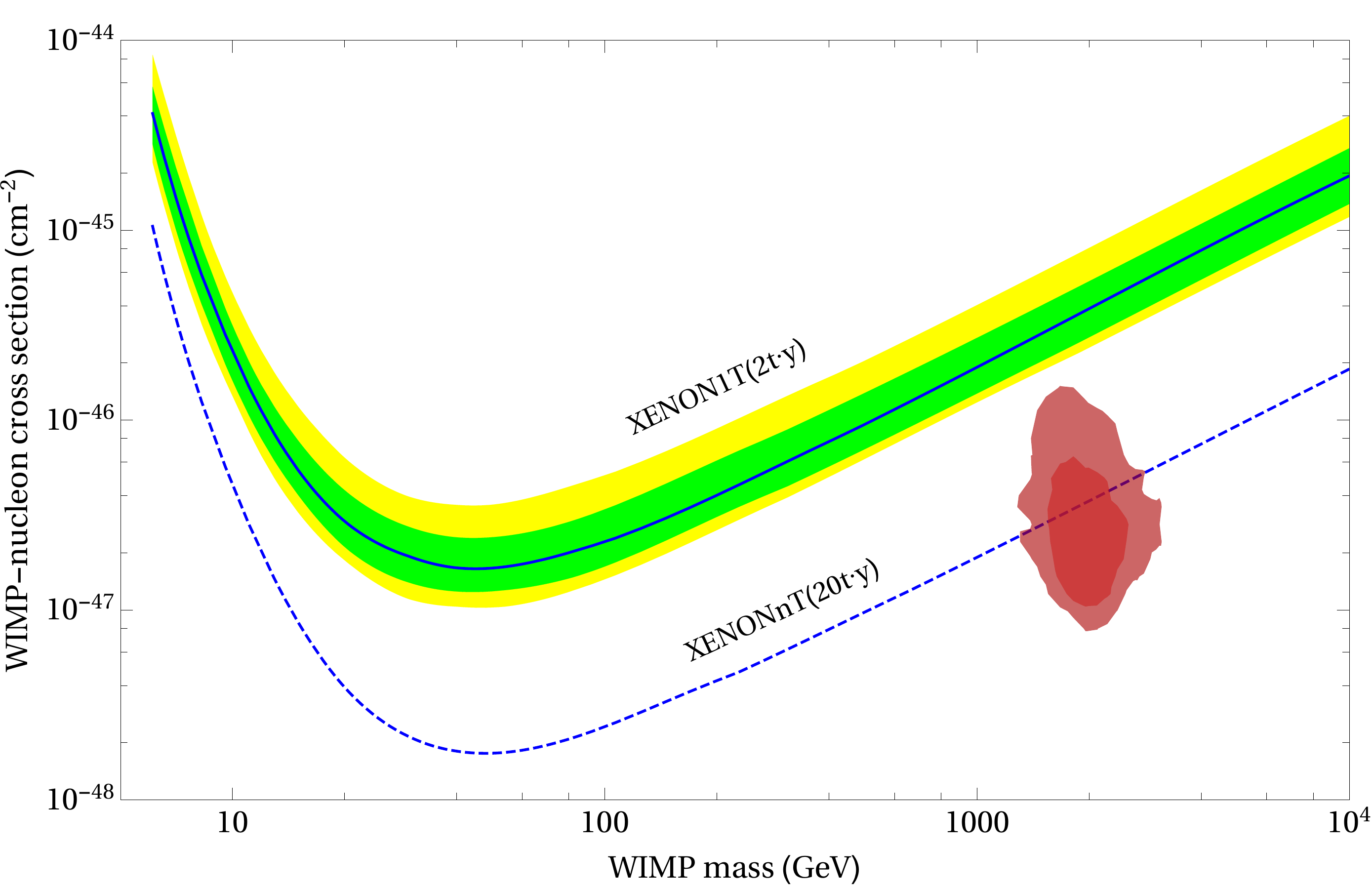}
\end{center}
\caption{XENON1T and XENONnT sensitivities to spin-independent WIMP-nucleon interaction as a function of the WIMP mass, shown as solid and dashed blue line. The green and yellow bands represent the $1\sigma$ and $2\sigma$ sensitivities of XENON1T \cite{Aprile:2015uzo} and the dark and light red regions are the $1\sigma$ and $2\sigma$ HPD regions of the MCMC analysis of our model. The WIMP is the LSP, which is the lightest neutralino.}
\label{fig:WimpNucleon}
\end{figure}

\begin{figure}
\begin{center}
\includegraphics[width=0.5\textwidth]{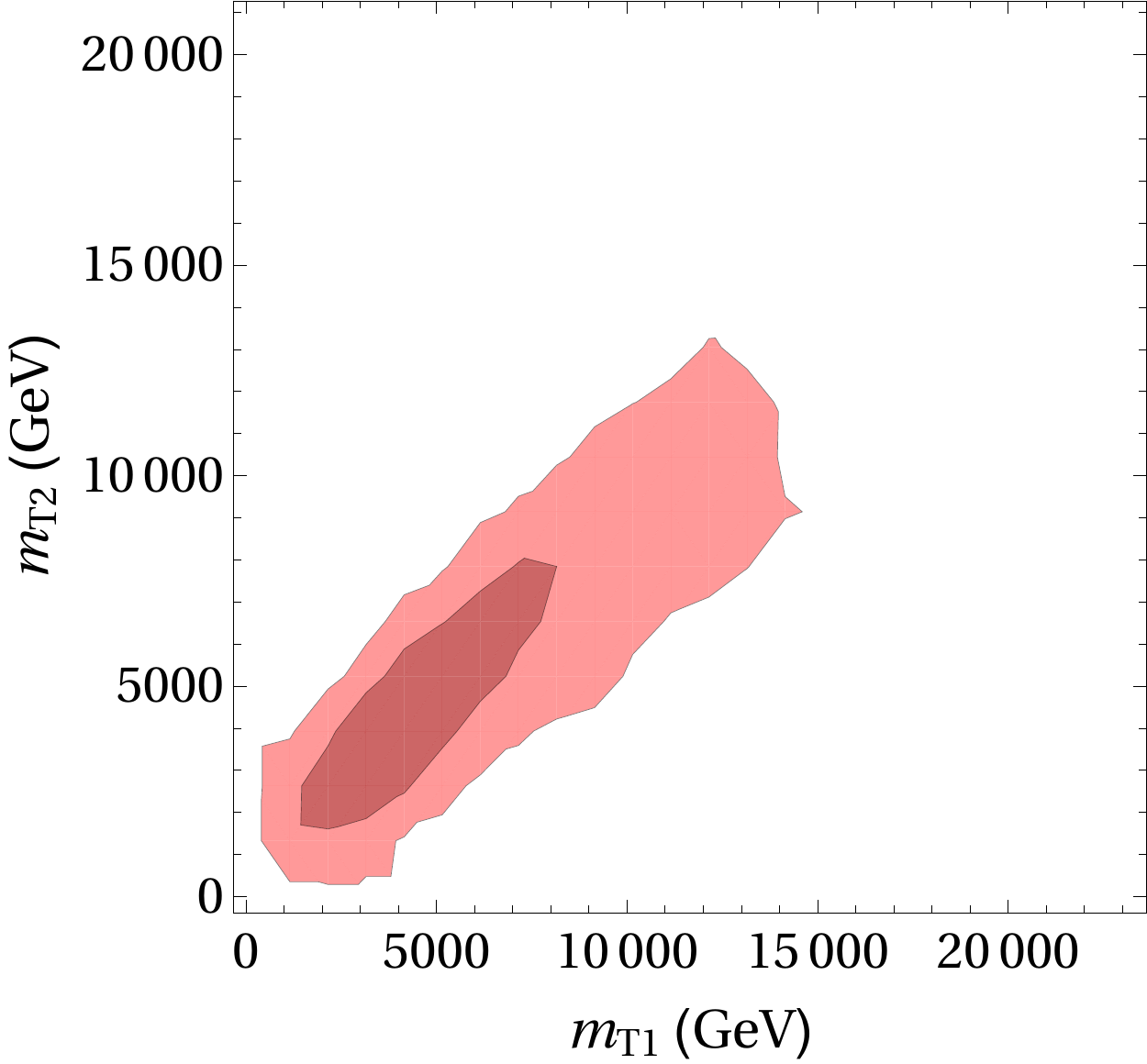}
\end{center}
\caption{Correlation between the soft mass parameters $m_{T1}$ and $m_{T2}$ defined in Eq.~(\ref{eq:SoftScalarMassMatrices_1}). The dark and the light red regions represent the $1\sigma$ and the $2\sigma$ HPD regions of the MCMC analysis, respectively.}
\label{fig:SoftMassesT1T2}
\end{figure}

\subsubsection{Discussion of the results \& testability}
Although in the MCMC analysis $30$ parameters are fitted to $30$ measured observables, we still make predictions for some of these observables, as can be see from Table~\ref{tab:BestFitParameters} and as will be discussed below. Furthermore, we also make predictions for yet unmeasured quantities which can be tested by future experiments:

\begin{itemize}
\item Since in our model all parameters in the soft terms are determined by the fit to the experimental data, we can calculate the SUSY spectrum. We predict $7446^{+1521}_{-1171}\,\text{GeV}$ for the SUSY scale, which is the characteristic mass scale of the supersymmetric partners of the SM particles. The whole SUSY spectrum and the heavy MSSM Higgs boson masses with the $1\sigma$ HPD intervals are shown in Figure~\ref{fig:MassSpectrum}. The lightest supersymmetric particle (LSP) is always the neutralino $\tilde{\chi}^0_1$ and the next-to-lightest supersymmetric particle (NLSP) is either a chargino, a stau or a sneutrino. The predicted HPD intervals for the sparticle masses are within the reach of possible future 100 TeV $pp$ colliders (see e.g.\ \cite{Golling:2016gvc}).

\item Connected to the above, in the dark matter sector we predict the correlation between the WIMP mass and the WIMP-nucleon cross section. Since the WIMP is the neutralino, Figure~\ref{fig:WimpNucleon} shows the $1\sigma$ and $2\sigma$ HPD regions of the MCMC analysis as well as the sensitivities of the XENON1T and the XENONnT experiments \cite{Aprile:2015uzo}. It turns out that our predictions lie beyond the range of XENON1T but there is an overlap with the sensitivity of XENONnT. 

\item In a small angle approximation of the charged lepton and the down-type Yukawa matrices one finds, in leading order, a simple relation between the ratio of the muon and electron Yukawa couplings, the ratio of the strange- and down-quark Yukawa couplings and the Clebsch-Gordan coefficients $c_{ij}$ (see Eq.~(\ref{eq:YukawaMatrices})) \cite{Antusch:2013jca}
\begin{align}
\label{eq:ClebschYukawa}
\frac{y_s}{y_d} \frac{y_e}{y_\mu} \approx \Big| \frac{c_{12}c_{21}}{c_{22}^2} \Big| \;.
\end{align}
From the experimental data (see Table~\ref{tab:BestFitObservables}) follows that the left hand side of Eq.~(\ref{eq:ClebschYukawa}) is given by $10.7^{+1.6}_{-0.7}$. In order to be in agreement with this value a suitable set of Clebsch-Gordan factors is mandatory. In our model the Clebsch-Gordon factors are given by $c_{12}=1$, $c_{21}=1$ and $c_{22}=-3$, what yields the value $9$ for the right hand side of Eq.~(\ref{eq:ClebschYukawa}). Since in our fit $y_e$ and $y_\mu$ are in good agreement with the experimental data, there is a deviation of $y_d$ and $y_s$ from their experimental values in order to compensate for the too small value delivered by the Clebsch-Gordon factors.

\item There is also a major contribution to the total $\chi^2$ from $\sin^2(\theta_{23}^{\text{PMNS}})$ and $\sin^2(\theta_{13}^{\text{PMNS}})$, whose values deviate from the experimental ones by more than $1\sigma$. Compared to the present experimental best-fit values, our model predicts somewhat smaller $\sin^2(\theta_{23}^{\text{PMNS}})$ and $\sin^2(\theta_{13}^{\text{PMNS}})$.

\item With vanishing 3-1 mixings in $Y_d$ and $Y_u$, the quark unitarity triangle angle $\alpha$ is given by the ``quark phase sum rule'' \cite{Antusch:2009hq}\footnote{Note, that in \cite{Antusch:2009hq} the LR convention is used in the Yukawa sector and not the RL convention as in our model.}
\begin{align}
\alpha \simeq \delta^d_{21} - \delta^u_{21} \;,
\end{align}
where $\delta^d_{21}$ and $\delta^u_{21}$ are the phases of the 2-1 mixings in the down- and in the up-type Yukawa coupling, respectively. When $\delta^d_{21} - \delta^u_{21}\simeq \pi/2$, as in our model (see Eq.~(\ref{eq:YukawaMatrices})), a realistic CKM CP phase $\delta^\mathrm{CKM}$ is induced \cite{Antusch:2009hq}. As shown in Table~\ref{tab:BestFitObservables}, the calculated value of $\delta^\mathrm{CKM}$ is indeed in good agreement with the experimental one, however the error bars in our model are much smaller than the ones of the experiment. This means we make an accurate prediction for the CKM CP phase, which can be tested by future experiments.

\item We find for the Dirac CP phase $\delta^{\text{PMNS}} = 4.0271^{+0.0035}_{-0.0028}$ in the PMNS matrix, what is in agreement with the $1\sigma$ range of the experimental data \cite{nufit}. Since the $3\sigma$ range of the experimental data is given by the whole interval between $0$ and $2\pi$, more precise measurements of $\delta^{\text{PMNS}}$ in the future have to show whether our model is excluded or not.

\item For the yet unmeasured Majorana phases of the PMNS matrix we predict $\varphi_1^{\text{PMNS}} = 0.3932^{+0.0129}_{-0.0117}$ and $\varphi_2^{\text{PMNS}} = 0.8005^{+0.0025}_{-0.0023}$.

\item The modulus of the $\mu$ parameter is determined by the requirement that the electroweak symmetry is broken, whereas the sign of $\mu$ has an influence on the ratio of the Yukawa couplings $y_\mu$ and $y_s$. Since the SUSY threshold corrections of $y_\mu$ and $y_s$ differ mostly by the $\tan\beta$-enhanced term including gluinos, which is proportional to $\mu$ (see e.g. \cite{Antusch:2015nwi}), the sign of $\mu$ has to be chosen in such a way that the ratio is in agreement with the experimental data. In our model we find $\mu = -5363^{+731}_{-752}\,\text{GeV}$. As a consequence of the negative sign of $\mu$ we predict in our model a smaller anomalous magnetic moment of the muon than in the SM, although this correction is small due to the high SUSY scale. Actual experimental data indicates that the correction of the SM value should go in the other direction, however there is still a big systematic uncertainty and, therefore, we did not include this observable in our analysis. On the other hand, a confirmation of the deviation from the SM value has the potential to exclude our model. 

\item Figure~\ref{fig:SoftMassesT1T2} shows the correlation between the parameters $m_{T1}$ and $m_{T2}$ in the soft mass matrix of the 10-plets (see Eq.~(\ref{eq:SoftScalarMassMatrices_1})). The plot indicates that it is favourable to have an universal value in these two entries of the soft mass matrix. The branching ratio of the flavour violating process $\mu \rightarrow e \gamma$ is highly sensitive to off-diagonal elements in the $2\times2$ soft mass matrix of the right-handed selectron and smuon in the SCKM basis. These off-diagonal entries are induced by non-universal soft masses in the flavour basis, since there is a non-zero mixing in the Yukawa matrix $Y_e$ (see Eq.~(\ref{eq:YukawaMatrices})) between the first and the second family. A mild correlation between $m_{T1}$ and $m_{T2}$ is sufficient to stay within the present bound on $Br( \mu \rightarrow e \gamma )$. On the other hand, a very strong suppression of $Br( \mu \rightarrow e \gamma )$ is not expected in our model.

\item For the branching ratios of the flavour violating processes $K^+ \rightarrow \pi^+\bar{\nu}\nu$, $B^0_S \rightarrow \mu^+\mu^-$ and $B \rightarrow \tau\nu$ the calculated values in the MCMC analysis lie already at the edge of the $1\sigma$ intervals of the experimental values (see Table~\ref{tab:BestFitObservables}). If in future experiments the error bars decrease further, our model can be tested. For the branching ratios of $K^0_L \rightarrow \pi^0\bar{\nu}\nu$ and $B^0_S \rightarrow e^+e^-$ we make precise predictions too, but the values lie far below the bounds from present experiments.
\end{itemize}

In summary, due to the added SUSY breaking sector the model can make various additional predictions for yet unmeasured quantities, such as the sparticle masses, dark matter properties and flavour violating processes, which increase the testability of the model.

\section{Summary and Conclusions}
We discussed how flavour GUT models can be combined with a SUSY breaking sector in the context of supergravity. We considered SUSY flavour GUT models where the flavour structure is generated when a family symmetry is spontaneously broken by vacuum expectation values of the scalar components of flavon superfields. In general there are $N_\mathrm{F}$ flavon superfields and $N_\mathrm{D}$ driving superfields in such a model. We showed that after a proper redefinition of the driving superfields one ends up with $N_\mathrm{F}$ superpotential contributions for driving the vevs of the flavon superfields and $N_\mathrm{D}-N_\mathrm{F}$ superpotential contributions suitable for SUSY breaking. For the case of one SUSY breaking field we explicitly constructed a SUSY breaking sector and showed that SUSY and family symmetry breaking can be combined in a consistent way.

A flavour model was constructed, following closely \cite{Antusch:2013wn}. The model is based on an SU(5) GUT symmetry, an $A_4$ family symmetry and a $\mathbb{Z}_4^\mathrm{R}$ symmetry, plus additional discrete shaping symmetries. The model breaks CP symmetry spontaneously and the lepton mixing is predicted by the CSD2 scheme \cite{Antusch:2011ic} plus additional charged lepton mixing. In the Yukawa sector the Clebsch factors are chosen in such a way that the approximate GUT relations $y_\tau = y_b$ and $y_\mu = -3y_s$ hold. We explicitly worked out the GUT matter sector of the model, including the full flavon and messenger sectors. This model was combined with a SUSY breaking sector containing one chiral superfield, along the lines of the discussion in the first part.

In order to investigate phenomenological aspects of the model we calculated the corresponding soft terms at the GUT scale, which emerge once SUSY is spontaneously broken, and determined the free parameters in the Yukawa sector, in the soft scalar masses and in the soft scalar trilinear couplings. Taking into account the RG evolution of the parameters between the GUT and the electroweak scale, as well as SUSY threshold corrections, we performed a Markov chain Monte Carlo fit to low energy observables, in particular Yukawa couplings, CKM and PMNS parameters, the SM-like EW Higgs boson mass, the dark matter relic density and flavour violating processes. Since all soft parameters are determined by this fit, the whole SUSY spectrum can be predicted. We find that for the analysed example model, the predicted highest posterior density intervals for the sparticle masses are within the reach of possible future 100 TeV $pp$ colliders (see e.g.\ \cite{Golling:2016gvc}).
Predictions were also made for the heavy MSSM Higgs boson masses as well as for the Dirac and Majorana phases in the PMNS matrix and for flavour violating processes.


\section*{Acknowledgements}
This work has been supported by the Swiss National Science Foundation. 

\appendix

\section{The superpotential}
\label{secApp:Superpotential}
In this appendix the superpotential of the flavour model, including the messenger fields (but without the SUSY breaking field), is presented. Since effective operators suppressed by the Planck scale are neglected in the flavour sector, only operators with at most dimension three are part of the superpotential. Integrating out the messenger fields, the effective operators as discussed before are obtained.
\\\\
The part of the superpotential which contains bilinear terms of the fields is only given by the mass terms for the messengers
\begin{align}
\label{eq:WrenMessengerMasses}
W^{\text{ren}}_\Lambda &= M_{\Gamma_i} \Gamma_i \bar{\Gamma}_i + M_{\Sigma_i} \Sigma_i \bar{\Sigma}_i + M_{\Omega_i} \Omega_i \bar{\Omega}_i + M_{\Xi_1} \Xi_1 \bar{\Xi}_1 \;.
\end{align}
The full list of the messengers and the corresponding representations and charges under the symmetries of the model is shown in Table~\ref{tab:MessengerFieldsList}. The masses in Eq.~(\ref{eq:WrenMessengerMasses}) are assumed to be bigger than the GUT scale such that the messengers can be integrated out to give the desired effective operators. To simplify the notation, before the messenger scale $\Lambda$ was written as a shorthand. It is related to the individual messenger masses and order one coefficients.
\\\\
The superpotential of the flavon sector is given by (where order one coefficients are dropped for the sake of readability)
\begin{align}
\begin{split}
W^{\text{ren}}_{\text{flavon}} &= O_{1;2} \theta_1 \theta_2 + O_{1;3} \theta_1 \theta_3 + O_{2;3} \theta_2 \theta_3 + O_{111;211} \theta_{111} \theta_{211} + O_{111;23} \theta_{111} \theta_{23} \\
& \quad + O_{23;211} \theta_{23} \theta_{211} + O_{2;102} \theta_2 \theta_{102} + O_{211;102} \theta_{211} \theta_{102} + O_{1;23} \theta_1 \theta_{23} \\
& \quad + A_1 \theta_1^2 + A_2 \theta_2^2 + A_3 \theta_3^2 + A_{111} \big( \theta_{111}^2 + \theta_{111} \rho_{111} + \theta_{111} \tilde{\rho}_{111} \big) \\
& \quad + P \Gamma_9 \xi_u + \bar{\Gamma}_9 \xi_u^2 + P \Gamma^2_8 + \bar{\Gamma}_8 \theta^2_2 + \bar{\Gamma}_8 \theta^{\prime 2}_2 + P \Gamma^2_7 + \bar{\Gamma}_7 \big( \theta^2_{111} + \rho^2_{111} + \tilde{\rho}^2_{111} \big) \\
& \quad + P \theta_{211} \Gamma_6 + \theta^2_{211} \bar{\Gamma}_6 + P \xi_2 \Gamma_5 + \xi^2_2 \bar{\Gamma}_5 + P \xi_1 \Gamma_4 + \xi^2_1 \bar{\Gamma}_4 + P \rho_{23} \Gamma_3 + \big( \theta^2_{23} + \rho^2_{23} \big) \bar{\Gamma}_3 \\
& \quad + P \rho_{102} \Gamma_2 + \big( \theta^2_{102} + \rho^2_{102} \big) \bar{\Gamma}_2 + P \theta^{\prime}_{102} \Gamma_1 + \theta^{\prime 2}_{102} \bar{\Gamma}_1 \;.
\end{split}
\end{align}
The first three lines are used to fix the flavon alignment in the vacuum as dicussed in chapter~4 in \cite{Antusch:2013wn}, while the last four lines are needed to fix the phases of the flavon vevs. A list of the driving fields is shown in Table~\ref{tab:DrivingFieldsList}.
\\\\
The dimension three operators including the matter and the Higgs fields have the form (again dropping order one coefficients)
\begin{align}
\begin{split}
W^{\text{ren}}_d &= T_3 \bar{H}_5 \bar{\Sigma}_3 + F \theta_3 \Sigma_3 + F \theta_{23} \Sigma_1 + T_2 \bar{\Sigma}_1 \bar{\Xi}_1 + \bar{H}^\prime_5 H_{24} \Xi_1 \\
& \quad + F \theta_{102} \Sigma_2 + T_2 \bar{\Sigma}_2 \bar{\Sigma}_6 + \theta_{102} \Sigma_6 \bar{\Sigma}_4 + H_{24} \bar{H}_5 \Sigma_4 \\
& \quad + T_1 \theta_2 \bar{\Omega}_4 + F \Omega_4 \bar{\Sigma}_5 + \theta_2 \Sigma_5 \bar{\Sigma}_4 \;, \\\\
\end{split} \\
\begin{split}
W^{\text{ren}}_u &= T_1 H_5 \Omega_3 + \xi_1 \Omega_2 \bar{\Omega}_3 + T_1 \xi_u \bar{\Omega}_2 + \Omega_2 \xi_u \bar{\Omega}_1 + T_2 \Gamma_4 \bar{\Omega}_1 \\
& \quad + \bar{\Gamma}_4 \xi^2_1 + T_2 H_5 \Omega_1 + T_3 \xi_1 \bar{\Omega}_1 + T^2_3 H_5 \;, \\\\
\end{split} \\
\begin{split}
W^{\text{ren}}_\nu &= \xi_1 N^2_1 + \xi_2 N^2_2 + N_1 H_5 \bar{\Sigma}_1 + N_2 H_5 \bar{\Sigma}_2 \;. \\
\end{split}
\end{align}
After integrating out the messenger fields one obtains the effective superpotential shown in Section~\ref{sec:SUSYBreakingMatterSuperpotential}.
\\\\
Beside the operators shown so far, there are some more dimension three operators in the superpotential which are allowed by the symmetries of the model
\begin{align}
\label{eq:WrenNeg}
\begin{split}
W^{\text{ren}}_{\text{neg}} &= T_1 \Gamma_9 \bar{\Omega}_1 + T_2 \Gamma_9 \bar{\Omega}_3 + \Gamma_9 \Omega_1 \bar{\Omega}_2 + \Gamma_4 \bar{\Omega}_2 \Omega_3 + \Gamma_1 \Sigma_4 \bar{\Sigma}_6 \\
& \quad + P A_2^2 + P A_{111}^2 + A_{111} \tilde{\rho}_{111} \theta_{111} + O_{211;211}^3\\
& \quad + \bar{\Gamma}^3_1 + \bar{\Gamma}^3_2 + \bar{\Gamma}^3_3 + \bar{\Gamma}^3_4 + \bar{\Gamma}^3_5 + \bar{\Gamma}^3_6 + P \bar{\Gamma}_7^2 + \bar{\Gamma}_7 \rho_{111} \tilde{\rho}_{111} + P \bar{\Gamma}_8^2 + \bar{\Gamma}^3_9 \;.
\end{split}
\end{align}
The first two give a contribution to the effective operator $T_1 T_2 H_5 \xi^2_u$ in the up-type quark sector, which is of the same order of magnitude as the one coming from the terms in $W^{\text{ren}}_u$. The other operators in Eq.~(\ref{eq:WrenNeg}) induce, after integrating out the messenger fields, operators of at least dimension seven, what gives only small corrections (since the effective operators emerging from the other parts of the superpotential have at most dimension six).

\bibliography{references}

\begin{table}[hbtp]
\renewcommand{\arraystretch}{1.3}
\begin{center}
\begin{tabular}{|rl|rrl|}
\hline
\multicolumn{2}{|c|}{Parameter} & \multicolumn{3}{c|}{MCMC}\\
\hline \hline
$\epsilon_2$ & in $10^{-7}$ & & $7.206$ & $^{+0.427}_{-0.414}$\\
$\epsilon_{102}$ & in $10^{-7}$ & & $-6.280$ & $^{+0.312}_{-0.450}$\\
$\epsilon_{23}$ & in $10^{-6}$ & & $-3.240$ & $^{+0.186}_{-0.192}$\\
$\epsilon_3$ & in $10^{-4}$ & & $-1.858$ & $^{+0.070}_{-0.090}$\\

\multirow{2}{*}{$a_u$} & \multirow{2}{*}{in $10^{-5}$} & \multirow{2}{*}{$\Bigg\{$} & $-1.182$ & $^{+0.127}_{-0.118}$\\ 
& & & $-0.741$ & $^{+0.117}_{-0.109}$\\

$b_u$ & in $10^{-4}$ & & $1.102$ & $^{+0.054}_{-0.060}$\\
$c_u$ & in $10^{-3}$ & & $-1.069$ & $^{+0.048}_{-0.046}$\\
$d_u$ & in $10^{-3}$ & & $9.403$ & $^{+0.811}_{-0.877}$\\
$e_u$ & in $10^{-1}$ & & $4.629$ & $^{+0.154}_{-0.143}$\\
$a$ & in $10^{-3}$ & & $-3.848$ & $^{+0.046}_{-0.044}$\\
$b$ & in $10^{-3}$ & & $-4.551$ & $^{+0.060}_{-0.047}$\\
$\tan\beta$ & & & $47.68$ & $^{+4.95}_{-3.11}$\\
\hline
$k_1$ & in $\text{GeV}$ & & $-8122$ & $^{+9686}_{-1.878}$\\
$k_2$ & in $\text{GeV}$ & & $-$ &\\
$k_3$ & in $\text{GeV}$ & & $8397$ & $^{+1603}_{-8233}$\\
$k_4$ & in $\text{GeV}$ & & $2$ & $^{+4201}_{-4745}$\\
$k_5$ & in $\text{GeV}$ & & $-$ &\\
$k_6$ & in $\text{GeV}$ & & $-$ &\\
$k_7$ & in $\text{GeV}$ & & $-$ &\\
$k_8$ & in $\text{GeV}$ & & $-$ &\\
$k_9$ & in $\text{GeV}$ & & $-3551$ & $^{+4062}_{-3679}$\\
$k_{10}$ & in $\text{GeV}$ & & $-$ &\\
$k_{11}$ & in $\text{GeV}$ & & $-$ &\\
\hline
$m_F$ & in $\text{GeV}$ & & $2595$ & $^{+1847}_{-1965}$\\
$m_{T_1}$ & in $\text{GeV}$ & & $3454$ & $^{+3299}_{-1908}$\\
$m_{T_2}$ & in $\text{GeV}$ & & $3314$ & $^{+2842}_{-1644}$\\
$m_{T_3}$ & in $\text{GeV}$ & & $4564$ & $^{+1930}_{-1556}$\\
$m_{H_u}$ & in $\text{GeV}$ & & $3294$ & $^{+2043}_{-2500}$\\
$m_{H_d}$ & in $\text{GeV}$ & & $2297$ & $^{+1581}_{-2171}$\\
\hline
$m_\lambda$ & in $\text{GeV}$ & & $4193$ & $^{+638}_{-761}$\\
\hline
\end{tabular}
\end{center}
\renewcommand{\arraystretch}{1}
\caption{The $1\sigma$ HPD intervals combined with the mode values of the parameters of the MCMC analysis. The ``$-$'' indicates that the parameter is uniformly distributed within the prior. The moduli of all $k_i$ are restricted to values smaller than $10\,\text{TeV}$. The two values of $a_u$ correspond to the two solutions of the (leading order) equation $y_u \approx |(Y_u)_{11} - (Y_u)^2_{12}/(Y_u)_{22}|$, where $(Y_u)_{11}=a_u$.}
\label{tab:BestFitParameters}
\end{table}
\begin{table}[hbtp]
\renewcommand{\arraystretch}{1.3}
\begin{center}
\begin{tabular}{|rl|rl|rl|}
\hline
\multicolumn{2}{|c|}{Observable} & \multicolumn{2}{c|}{Experiment} & \multicolumn{2}{c|}{MCMC}\\
\hline \hline
$y_u$ & in $10^{-6}$ & $7.4$ & $^{+1.5}_{-3.0}$ & $7.0$ & $^{+1.8}_{-2.7}$\\
$y_c$ & in $10^{-3}$ & $3.60$ & $\pm0.11$ & $3.61$ & $^{+0.10}_{-0.12}$\\
$y_t$ & & $0.9861$ & $^{+0.0086}_{-0.0087}$ & $0.9883$ & $^{+0.0078}_{-0.0097}$\\
\hline
$y_d$ & in $10^{-5}$ & $1.58$ & $^{+0.23}_{-0.10}$ & $1.44$ & $^{+0.05}_{-0.05}$\\
$y_s$ & in $10^{-4}$ & $3.12$ & $^{+0.17}_{-0.16}$ & $3.55$ & $^{+0.12}_{-0.10}$\\
$y_b$ & in $10^{-2}$ & $1.639$ & $\pm0.015$ & $1.636$ & $^{+0.016}_{-0.014}$\\
\hline
$y_e$ & in $10^{-6}$ & $2.795$ & $\pm1\%$ & $2.808$ & $^{+0.027}_{-0.029}$\\
$y_\mu$ & in $10^{-4}$ & $5.900$ & $\pm1\%$ & $5.857$ & $^{+0.054}_{-0.064}$\\
$y_\tau$ & in $10^{-2}$ & $1.003$ & $\pm1\%$ & $1.003$ & $^{+0.009}_{-0.010}$\\
\hline
$\theta^{\text{CKM}}_{12}$ & in $10^{-1}$ & $2.2735$ & $\pm0.0072$ & $2.2729$ & $^{+0.0072}_{-0.0078}$\\
$\theta^{\text{CKM}}_{23}$ & in $10^{-2}$ & $4.208$ & $\pm0.064$ & $4.211$ & $^{+0.057}_{-0.073}$\\
$\theta^{\text{CKM}}_{13}$ & in $10^{-3}$ & $3.64$ & $\pm0.13$ & $3.676$ & $^{+0.116}_{-0.142}$\\
$\delta^{\text{CKM}}$ & & $1.208$ & $\pm0.054$ & $1.207$ & $^{+0.015}_{-0.014}$\\
\hline
$\sin^2(\theta_{12}^{\text{PMNS}})$ & in $10^{-1}$ & $3.08$ & $^{+0.13}_{-0.12}$ & $3.08$ & $^{+0.02}_{-0.02}$\\
$\sin^2(\theta_{23}^{\text{PMNS}})$ & in $10^{-1}$ & $4.51$ & $^{+0.38}_{-0.25}$ & $4.11$ & $^{+0.06}_{-0.05}$\\
$\sin^2(\theta_{13}^{\text{PMNS}})$ & in $10^{-2}$ & $2.19$ & $\pm0.10$ & $2.03$ & $^{+0.05}_{-0.04}$\\
$\Delta m^2_\mathrm{sol}$ & in $10^{-5}$ & $7.49$ & $^{+0.19}_{-0.17}$ & $7.671$ & $^{+0.174}_{-0.173}$\\
$\Delta m^2_\mathrm{atm}$ & in $10^{-3}$ & $2.477$ & $\pm0.042$ & $2.505$ & $^{+0.036}_{-0.042}$\\
\hline
$m_h$ & in $\text{GeV}$ & $125.6$ & $\pm3.0$ & $125.7$ & $^{+0.9}_{-0.8}$\\
$\Omega$ & in $10^{-1}$ & $1.186$ & $\pm0.020^*$ & $1.219$ & $^{+0.219}_{-0.193}$\\
\hline
$\mu \rightarrow e\gamma$ & in $10^{-13}$ & $\le 5.7$ & & $\le 1.9$ &\\
$\tau \rightarrow e\gamma$ & in $10^{-16}$ & $\le 3.3\cdot 10^8$ & & $\le 9.8$ &\\
$\tau \rightarrow \mu\gamma$ & in $10^{-13}$ & $\le 4.4\cdot 10^5$ & & $\le 4.3$ &\\
$K^0_L \rightarrow \pi^0\bar{\nu}\nu$ & in $10^{-11}$ & $\le 2.6\cdot 10^3$ & & $2.839$ & $^{+0.007}_{-0.011}$\\
$K^+ \rightarrow \pi^+\bar{\nu}\nu$ & in $10^{-10}$ & $1.73$ & $^{+1.15}_{-1.05}$ & $0.7798$ & $^{+0.0013}_{-0.0021}$\\
$B^0_S \rightarrow e^+e^-$ & in $10^{-14}$ & $\le 2.8\cdot 10^7$ & & $4.97$ & $^{+0.55}_{-0.74}$\\
$B^0_S \rightarrow \mu^+\mu^-$ & in $10^{-9}$ & $3.1$ & $\pm0.7$ & $2.12$ & $^{+0.24}_{-0.32}$\\
$B^0_S \rightarrow e^\pm\mu^\mp$ & in $10^{-20}$ & $\le 1.1\cdot 10^{12}$ & & $\le 1.7$ &\\
$B \rightarrow \tau\nu$ & in $10^{-4}$ & $1.14$ & $\pm0.27$ & $0.886$ & $^{+0.005}_{-0.006}$\\
$\epsilon_K^{\text{exp}}/\epsilon^{\text{SM}}_K$ & & $1.09$ & $\pm0.16$ & $1.01$ & $^{+0.10}_{-0.10}$\\
\hline
\end{tabular}
\end{center}
\renewcommand{\arraystretch}{1}
\caption{Experimental values of the observables and the $1\sigma$ HPD intervals combined with the mode values of the observables of the MCMC analysis. The Yukawa couplings, the CKM and PMNS parameters, and the neutrino masses are given at $M_Z = 91.2\,\text{GeV}$. Note that although the Yukawa couplings of the charged leptons are measured far more precise than listed, we set an $1\%$ uncertainty for the experimental values, what is roughly in accordance with the accuracy of the running used here. The same holds true for the Higgs mass, where the uncertainty of the theoretical calculation of about $\pm 3\,\text{GeV}$ is much bigger than the experimental one. $^*$In the MCMC analysis ten times the listed error is used.}
\label{tab:BestFitObservables}
\end{table}

\begin{table}[hbtp]
\renewcommand{\arraystretch}{1.3}
\begin{center}
\begin{tabular}{|c||ccc|cc|cccc|}
\hline
 & $\mathrm{SU(5)}$ & $A_4$ & $\mathbb{Z}_4^\mathrm{R}$ & $\mathbb{Z}_4^{(a)}$ & $\mathbb{Z}_4^{(b)}$ & $\mathbb{Z}_3^{(a)}$ & $\mathbb{Z}_3^{(b)}$ & $\mathbb{Z}_3^{(c)}$ & $\mathbb{Z}_3^{(d)}$ \\
\hline \hline
\multicolumn{10}{|l|}{Matter fields}\\
\hline \hline
$F$ & $\overline{\pmb{5}}$ & $\pmb{3}$ & $1$ & $.$ & $.$ & $.$ & $.$ & $1$ & $2$\\
$T_1$ & $\pmb{10}$ & $.$ & $1$ & $3$ & $3$ & $1$ & $1$ & $.$ & $.$\\
$T_2$ & $\pmb{10}$ & $.$ & $1$ & $3$ & $3$ & $2$ & $1$ & $2$ & $.$\\
$T_3$ & $\pmb{10}$ & $.$ & $1$ & $3$ & $3$ & $.$ & $.$ & $2$ & $.$\\
$N_1$ & $\pmb{1}$ & $.$ & $1$ & $.$ & $2$ & $1$ & $2$ & $.$ & $.$\\
$N_2$ & $\pmb{1}$ & $.$ & $1$ & $2$ & $2$ & $2$ & $.$ & $2$ & $.$\\
\hline \hline
\multicolumn{10}{|l|}{Higgs fields}\\
\hline \hline
$H_5$ & $\pmb{5}$ & $.$ & $.$ & $2$ & $2$ & $.$ & $.$ & $2$ & $.$\\
$\bar{H}_5$ & $\overline{\pmb{5}}$ & $.$ & $.$ & $.$ & $.$ & $.$ & $.$ & $.$ & $.$\\
$\bar{H}_5^\prime$ & $\overline{\pmb{5}}$ & $.$ & $.$ & $2$ & $.$ & $.$ & $2$ & $1$ & $2$\\
$H_{45}$ & $\pmb{45}$ & $.$ & $.$ & $.$ & $2$ & $.$ & $1$ & $1$ & $1$\\
$H_{24}$ & $\pmb{24}$ & $.$ & $.$ & $1$ & $1$ & $2$ & $2$ & $2$ & $1$\\
$S$ & $.$ & $.$ & $2$ & $2$ & $2$ & $.$ & $.$ & $1$ & $.$\\
\hline \hline
\multicolumn{10}{|l|}{Flavon fields}\\
\hline \hline
$\theta_{102}$ & $.$ & $\pmb{3}$ & $.$ & $.$ & $.$ & $1$ & $.$ & $1$ & $1$\\
$\theta_{23}$ & $.$ & $\pmb{3}$ & $.$ & $2$ & $.$ & $2$ & $1$ & $.$ & $1$\\
$\theta_1$ & $.$ & $\pmb{3}$ & $.$ & $1$ & $3$ & $1$ & $.$ & $.$ & $1$\\
$\theta_2$ & $.$ & $\pmb{3}$ & $.$ & $.$ & $3$ & $.$ & $.$ & $.$ & $.$\\
$\theta_3$ & $.$ & $\pmb{3}$ & $.$ & $1$ & $1$ & $.$ & $.$ & $.$ & $1$\\
$\theta_{111}$ & $.$ & $\pmb{3}$ & $.$ & $3$ & $3$ & $.$ & $.$ & $.$ & $.$\\
$\theta_{211}$ & $.$ & $\pmb{3}$ & $.$ & $.$ & $.$ & $2$ & $1$ & $1$ & $.$\\
$\xi_u$ & $.$ & $.$ & $.$ & $.$ & $.$ & $.$ & $2$ & $1$ & $.$\\
$\xi_1$ & $.$ & $.$ & $.$ & $.$ & $.$ & $1$ & $2$ & $.$ & $.$\\
$\xi_2$ & $.$ & $.$ & $.$ & $.$ & $.$ & $2$ & $.$ & $2$ & $.$\\
$\theta_2^\prime$ & $.$ & $.$ & $.$ & $.$ & $1$ & $.$ & $.$ & $.$ & $.$\\
$\theta_{102}^\prime$ & $.$ & $.$ & $.$ & $.$ & $.$ & $1$ & $.$ & $.$ & $2$\\
$\rho_{111}$ & $.$ & $.$ & $.$ & $3$ & $3$ & $.$ & $.$ & $.$ & $.$\\
$\tilde{\rho}_{111}$ & $.$ & $.$ & $.$ & $3$ & $3$ & $.$ & $.$ & $.$ & $.$\\
$\rho_{23}$ & $.$ & $.$ & $.$ & $.$ & $.$ & $2$ & $1$ & $.$ & $1$\\
$\rho_{102}$ & $.$ & $.$ & $.$ & $.$ & $.$ & $1$ & $.$ & $1$ & $1$\\
\hline
\end{tabular}
\end{center}
\renewcommand{\arraystretch}{1}
\caption{The matter, Higgs and flavon field content of our model. A dot means that the field is an invariant singlet under the respective symmetry.}
\label{tab:MatterHiggsFlavonFieldsList}
\end{table}

\begin{table}[hbtp]
\renewcommand{\arraystretch}{1.3}
\begin{center}
\begin{tabular}{|c||ccc|cc|cccc|}
\hline
 & $\mathrm{SU(5)}$ & $A_4$ & $\mathbb{Z}_4^\mathrm{R}$ & $\mathbb{Z}_4^{(a)}$ & $\mathbb{Z}_4^{(b)}$ & $\mathbb{Z}_3^{(a)}$ & $\mathbb{Z}_3^{(b)}$ & $\mathbb{Z}_3^{(c)}$ & $\mathbb{Z}_3^{(d)}$ \\
\hline \hline
\multicolumn{10}{|l|}{Driving fields}\\
\hline \hline
$O_{1;2}$ & $.$ & $.$ & $2$ & $3$ & $2$ & $2$ & $.$ & $.$ & $2$\\
$O_{1;3}$ & $.$ & $.$ & $2$ & $2$ & $.$ & $2$ & $.$ & $.$ & $1$\\
$O_{2;3}$ & $.$ & $.$ & $2$ & $3$ & $.$ & $.$ & $.$ & $.$ & $2$\\
$O_{111;211}$ & $.$ & $.$ & $2$ & $1$ & $1$ & $1$ & $2$ & $2$ & $.$\\
$O_{111;23}$ & $.$ & $.$ & $2$ & $3$ & $1$ & $1$ & $2$ & $.$ & $2$\\
$O_{23;211}$ & $.$ & $.$ & $2$ & $2$ & $.$ & $2$ & $1$ & $2$ & $2$\\
$O_{2;102}$ & $.$ & $.$ & $2$ & $.$ & $1$ & $2$ & $.$ & $2$ & $2$\\
$O_{211;102}$ & $.$ & $.$ & $2$ & $.$ & $.$ & $.$ & $2$ & $1$ & $2$\\
$O_{1;23}$ & $.$ & $.$ & $2$ & $1$ & $1$ & $.$ & $2$ & $.$ & $1$\\
$A_1$ & $.$ & $\pmb{3}$ & $2$ & $2$ & $2$ & $1$ & $.$ & $.$ & $1$\\
$A_2$ & $.$ & $\pmb{3}$ & $2$ & $.$ & $2$ & $.$ & $.$ & $.$ & $.$\\
$A_3$ & $.$ & $\pmb{3}$ & $2$ & $2$ & $2$ & $.$ & $.$ & $.$ & $1$\\
$A_{111}$ & $.$ & $\pmb{3}$ & $2$ & $2$ & $2$ & $.$ & $.$ & $.$ & $.$\\
$P$ & $.$ & $.$ & $2$ & $.$ & $.$ & $.$ & $.$ & $.$ & $.$\\
\hline
\end{tabular}
\end{center}
\renewcommand{\arraystretch}{1}
\caption{The driving field content of our model. A dot means that the field is an invariant singlet under the respective symmetry. Note, that only one $P$ field is shown here. Indeed one has to introduce as many $P$ fields as operators to fix the phases of the flavon fields, as described in Section~\ref{sec:DrivingFields}.}
\label{tab:DrivingFieldsList}
\end{table}

\begin{table}[hbtp]
\renewcommand{\arraystretch}{1.3}
\begin{center}
\begin{tabular}{|c||ccc|cc|cccc|}
\hline
 & $\mathrm{SU(5)}$ & $A_4$ & $\mathbb{Z}_4^\mathrm{R}$ & $\mathbb{Z}_4^{(a)}$ & $\mathbb{Z}_4^{(b)}$ & $\mathbb{Z}_3^{(a)}$ & $\mathbb{Z}_3^{(b)}$ & $\mathbb{Z}_3^{(c)}$ & $\mathbb{Z}_3^{(d)}$ \\
\hline \hline
\multicolumn{10}{|l|}{Messenger fields}\\
\hline \hline
$\Gamma_1,\;\bar{\Gamma}_1$ & $.$ & $.$ & $0,\;2$ & $.$ & $.$ & $2,\;1$ & $.$ & $.$ & $1,\;2$\\
$\Gamma_2,\;\bar{\Gamma}_2$ & $.$ & $.$ & $0,\;2$ & $.$ & $.$ & $2,\;1$ & $.$ & $2,\;1$ & $2,\;1$\\
$\Gamma_3,\;\bar{\Gamma}_3$ & $.$ & $.$ & $0,\;2$ & $.$ & $.$ & $1,\;2$ & $2,\;1$ & $.$ & $2,\;1$\\
$\Gamma_4,\;\bar{\Gamma}_4$ & $.$ & $.$ & $0,\;2$ & $.$ & $.$ & $2,\;1$ & $1,\;2$ & $.$ & $.$\\
$\Gamma_5,\;\bar{\Gamma}_5$ & $.$ & $.$ & $0,\;2$ & $.$ & $.$ & $1,\;2$ & $.$ & $1,\;2$ & $.$\\
$\Gamma_6,\;\bar{\Gamma}_6$ & $.$ & $\pmb{3},\;\pmb{3}$ & $0,\;2$ & $.$ & $.$ & $1,\;2$ & $2,\;1$ & $2,\;1$ & $.$\\
$\Gamma_7,\;\bar{\Gamma}_7$ & $.$ & $.$ & $0,\;2$ & $2,\;2$ & $2,\;2$ & $.$ & $.$ & $.$ & $.$\\
$\Gamma_8,\;\bar{\Gamma}_8$ & $.$ & $.$ & $0,\;2$ & $.$ & $2,\;2$ & $.$ & $.$ & $.$ & $.$\\
$\Gamma_9,\;\bar{\Gamma}_9$ & $.$ & $.$ & $0,\;2$ & $.$ & $.$ & $.$ & $1,\;2$ & $2,\;1$ & $.$\\
\hline
$\Sigma_1,\;\bar{\Sigma}_1$ & $\pmb{5},\;\overline{\pmb{5}}$ & $.$ & $1,\;1$ & $2,\;2$ & $.$ & $1,\;2$ & $2,\;1$ & $2,\;1$ & $.$\\
$\Sigma_2,\;\bar{\Sigma}_2$ & $\pmb{5},\;\overline{\pmb{5}}$ & $.$ & $1,\;1$ & $.$ & $.$ & $2,\;1$ & $.$ & $1,\;2$ & $.$\\
$\Sigma_3,\;\bar{\Sigma}_3$ & $\pmb{5},\;\overline{\pmb{5}}$ & $.$ & $1,\;1$ & $3,\;1$ & $3,\;1$ & $.$ & $.$ & $2,\;1$ & $.$\\
$\Sigma_4,\;\bar{\Sigma}_4$ & $\pmb{5},\;\overline{\pmb{5}}$ & $.$ & $2,\;0$ & $3,\;1$ & $3,\;1$ & $1,\;2$ & $1,\;2$ & $1,\;2$ & $2,\;1$\\
$\Sigma_5,\;\bar{\Sigma}_5$ & $\pmb{5},\;\overline{\pmb{5}}$ & $.$ & $2,\;0$ & $3,\;1$ & $2,\;2$ & $1,\;2$ & $1,\;2$ & $1,\;2$ & $2,\;1$\\
$\Sigma_6,\;\bar{\Sigma}_6$ & $\pmb{5},\;\overline{\pmb{5}}$ & $.$ & $2,\;0$ & $3,\;1$ & $3,\;1$ & $.$ & $1,\;2$ & $1,\;2$ & $.$\\
\hline
$\Omega_1,\;\bar{\Omega}_1$ & $\pmb{10},\;\overline{\pmb{10}}$ & $.$ & $1,\;1$ & $3,\;1$ & $3,\;1$ & $1,\;2$ & $2,\;1$ & $2,\;1$ & $.$\\
$\Omega_2,\;\bar{\Omega}_2$ & $\pmb{10},\;\overline{\pmb{10}}$ & $.$ & $1,\;1$ & $3,\;1$ & $3,\;1$ & $1,\;2$ & $.$ & $1,\;2$ & $.$\\
$\Omega_3,\;\bar{\Omega}_3$ & $\pmb{10},\;\overline{\pmb{10}}$ & $.$ & $1,\;1$ & $3,\;1$ & $3,\;1$ & $2,\;1$ & $2,\;1$ & $1,\;2$ & $.$\\
$\Omega_4,\;\bar{\Omega}_4$ & $\pmb{10},\;\overline{\pmb{10}}$ & $\pmb{3},\;\pmb{3}$ & $1,\;1$ & $3,\;1$ & $2,\;2$ & $1,\;2$ & $1,\;2$ & $.$ & $.$\\
\hline
$\Xi_1,\;\bar{\Xi}_1$ & $\pmb{45},\;\overline{\pmb{45}}$ & $.$ & $2,\;0$ & $1,\;3$ & $3,\;1$ & $1,\;2$ & $2,\;1$ & $.$ & $.$\\
\hline
\end{tabular}
\end{center}
\renewcommand{\arraystretch}{1}
\caption{The messenger field content of our model. A dot means that the field is an invariant singlet under the respective symmetry.}
\label{tab:MessengerFieldsList}
\end{table}

\end{document}